\documentclass[article]{jss}

\usepackage{thumbpdf,lmodern}
\usepackage{amsmath,amsthm,amssymb,amscd}
\usepackage{bm}
\usepackage{graphicx, subfig}
\usepackage{enumitem, calc}

\usepackage{framed}

\author{Sheng Dai\\Aalto University School of Business \And Yu-Hsueh Fang \\National Cheng Kung University
\AND Chia-Yen Lee\\National Taiwan University \And Timo Kuosmanen\\Aalto University School of Business}
\Plainauthor{Sheng Dai, Yu-Hsueh Fang, Chia-Yen Lee, and Timo Kuosmanen}

\title{\vspace{-2.3cm}\pkg{pyStoNED}: A \proglang{Python} Package for Convex Regression and Frontier Estimation}
\Plaintitle{pyStoNED: A Python Package for Convex Regression and Frontier Estimation}
\Shorttitle{ }

\Abstract{
Shape-constrained nonparametric regression is a growing area in econometrics, statistics, operations research, machine learning and related fields. In the field of productivity and efficiency analysis, recent developments in the multivariate convex regression and related techniques such as convex quantile regression and convex expectile regression have bridged the long-standing gap between the conventional deterministic-nonparametric and stochastic-parametric methods. Unfortunately, the heavy computational burden and the lack of powerful, reliable, and fully open access computational package has slowed down the diffusion of these advanced estimation techniques to the empirical practice. The purpose of the \proglang{Python} package \pkg{pyStoNED} is to address this challenge by providing a freely available and user-friendly tool for the multivariate convex regression, convex quantile regression, convex expectile regression, isotonic regression, stochastic nonparametric envelopment of data, and related methods. This paper presents a tutorial of the \pkg{pyStoNED} package and illustrates its application, focusing on the estimation of frontier cost and production functions.
}

\Keywords{multivariate convex regression, nonparametric least squares, frontier estimation, efficiency analysis, stochastic noise, \proglang{Python}}
\Plainkeywords{efficiency analysis, frontier estimation, multivariate convex regression, nonparametric least squares, stochastic noise, python}

\Address{
  Sheng Dai, Timo Kuosmanen\\
  Department of Information and Service Management\\
  Aalto University School of Business\\
  02150 Espoo, Finland\\
  E-mail: \email{sheng.dai@aalto.fi, timo.kuosmanen@aalto.fi}\\
  \\
  Yu-Hsueh Fang \\
  Institute of Manufacturing Information and Systems\\
  National Cheng Kung University\\
  Tainan, Taiwan \\
  Email: \email{p96094066@gs.ncku.edu.tw}\\
  \\
  Chia-Yen Lee \\
  Department of Information Management\\
  National Taiwan University \\
  Taipei City 106, Taiwan \\
  Email: \email{chiayenlee@ntu.edu.tw}
}

\begin{document}

\section{Introduction} \label{sec:intro}

Early contributions to nonparametric regression by \citet{Hildreth1954}, \citet{Brunk1955}, and \citet{Grenander1956} built exclusively on the convexity and monotonicity constraints of the regression function. However, extending these approaches from the univariate setting to the more general multivariate regression proved a vexing challenge. Since the development of an explicit piece-wise linear characterization of the multivariate convex nonparametric least squares (CNLS) by \citet{Kuosmanen2008}, convex regression has attracted growing interest in econometrics, statistics, operations research, machine learning and related fields (e.g., \citeauthor{Magnani2009}, \citeyear{Magnani2009}; \citeauthor{Seijo2011}, \citeyear{Seijo2011}; \citeauthor{Lim2012}, \citeyear{Lim2012}; \citeauthor{Hannah2013}, \citeyear{Hannah2013}; \citeauthor{Mazumder2019}, \citeyear{Mazumder2019}; \citeauthor{Bertsimas2020}, \citeyear{Bertsimas2020}). The recent study by \citet{Yagi2018} applies insights from convex regression to impose shape constraints on a local polynomial kernel estimator.

Convexity and monotonicity constraints are particularly relevant in the microeconomic applications where the duality theory of production and consumption directly implies certain monotonicity and convexity/concavity properties for many functions of interest (e.g., \citeauthor{Afriat1967}, \citeyear{Afriat1967}, \citeyear{Afriat1972}; \citeauthor{Varian1982}, \citeyear{Varian1982}, \citeyear{Varian1984}). For example, the cost function of the firm must be monotonic increasing and convex with respect to the input prices. Similar to the fact that a density function must be non-negative and its definite integral is equal to one, the cost function must satisfy the monotonicity and convexity properties implied by the theory, otherwise it is not really a cost function at all. The recent developments in the convex regression enable researchers to impose the concavity or convexity constraints implied by the theory to estimate the functions of interest without any parametric functional form assumptions. 

In recent years, convex regression and related techniques are increasingly utilized in the estimation of frontier cost and production functions in the field of productivity and efficiency analysis, a multidisciplinary field that is widely applied in such areas as agriculture, banking, education, environment, health care, energy, manufacturing, transportation, and utilities (e.g., \citeauthor{Kuosmanen2015d}, \citeyear{Kuosmanen2015d}; \citeauthor{Ray2015}, \citeyear{Ray2015}). Traditionally, this field was divided between two competing paradigms: Data Envelopment Analysis (DEA) (\citeauthor{Charnes1978}, \citeyear{Charnes1978}) and Stochastic Frontier Analysis (SFA) (\citeauthor{Aigner1977}, \citeyear{Aigner1977}; \citeauthor{Meeusen1977}, \citeyear{Meeusen1977}). DEA is a deterministic, fully nonparametric approach whereas SFA is a probabilistic, fully parametric approach. To bridge the gap between these two paradigms, Stochastic nonparametric envelopment of data (StoNED) (\citeauthor{Kuosmanen2006}, \citeyear{Kuosmanen2006}; \citeauthor{Kuosmanen2012c}, \citeyear{Kuosmanen2012c}) was proposed as a unified framework that combines virtues of DEA and SFA, encompassing both approaches as its restricted special cases. 

In practice, convex regression and StoNED are computationally demanding approaches, requiring a user to solve a mathematical programming problem subject to a large number of linear constraints. For example, the additive CNLS formulation by \citet{Kuosmanen2008} is a quadratic programming (QP) problem, whereas the multiplicative logarithmic formulation first considered by \citet{Kuosmanen2012c} requires solving a nonlinear programming (NLP) problem. Therefore, most empirical applications published thus far make use of commercial QP and NLP solvers, which can be coded using high-level mathematical computing languages such as \proglang{GAMS} (\citeauthor{GamsSoftware2013}, \citeyear{GamsSoftware2013}) or \proglang{MATLAB} (\citeauthor{MATLAB}, \citeyear{MATLAB}). \citet{Ray2015} present detailed examples of how to implement the most basic CNLS and StoNED formulations in \proglang{MATLAB} and \proglang{GAMS}. Recently, the \pkg{Benchmarking} package (\citeauthor{bogetoft2010}, \citeyear{bogetoft2010}) implemented in \proglang{R} (\citeauthor{R}, \citeyear{R}) includes a new function StoNED() to estimate the CNLS/StoNED model, however, the \proglang{R} implementation is currently restricted to the additive CNLS QP formulation model. 

The lack of a comprehensive, powerful, reliable, and fully open access computational package for the CNLS, StoNED, and related methods has slowed down the diffusion of these techniques to the empirical practice, which still heavily relies on the simple DEA and SFA techniques that either assume away noise or rely on restrictive functional form assumptions.\footnote{A plethora of computational tools, packages and resources are available for the conventional DEA and SFA. In the open access environment (\proglang{R}), one could use \pkg{Benchmarking} (\citeauthor{bogetoft2010}, \citeyear{bogetoft2010}), \pkg{FEAR} (\citeauthor{wilson2008}, \citeyear{wilson2008}), \pkg{rDEA} (\citeauthor{simm2020}, \citeyear{simm2020}), \pkg{nonparaeff} (\citeauthor{oh2013}, \citeyear{oh2013}), \pkg{npsf} (\citeauthor{oleg2020}, \citeyear{oleg2020}), and \pkg{frontiles} (\citeauthor{Daouia2020}, \citeyear{Daouia2020}), among many others, for solving different types of DEA models. For the SFA models, there exist \pkg{frontier}(\citeauthor{Tim2020}, \citeyear{Tim2020}), \pkg{sfa} (\citeauthor{ariane2014}, \citeyear{ariane2014}), and \pkg{sfaR} (\citeauthor{Dakpo2021}, \citeyear{Dakpo2021}). Other free packages such as \pkg{pyDEA} (\citeauthor{pyDEA}, \citeyear{pyDEA}) and \pkg{DataEnvelopmentAnalysis} (\citeauthor{deajl}, \citeyear{deajl}) are available to implement efficiency analysis. Further, there are also a plenty of DEA and SFA codes and packages for \proglang{MATLAB}, \proglang{GAMS}, and other high-level mathematical computing languages (e.g., \citeauthor{kalvelagen2002}, \citeyear{kalvelagen2002}; \citeauthor{ji2010}, \citeyear{ji2010}; \citeauthor{belotti2013}, \citeyear{belotti2013}; \citeauthor{Alvarez2020}, \citeyear{Alvarez2020}). 
}
To lower the barrier for applied researchers and practitioners to apply more advanced techniques that help to relax unnecessarily restrictive assumption, the \proglang{Python} package \pkg{pyStoNED} was first introduced in April 2020 to prove a freely available and user-friendly tool for the multivariate CNLS and StoNED methods. Its latest edition 0.5.2 also includes modules for the convex quantile regression, convex expectile regression, isotonic regression, and graphical illustration. It also facilitates efficiency measurement using the conventional DEA and free disposable hull (FDH) approaches. The \pkg{pyStoNED} package allows practitioners to estimate these models in an open access environment under a GPL-3.0 License. The project, including source code, internal data, notebook tutorials, and web documentation, is publicly available at GitHub.\footnote{
The \pkg{pyStoNED} package GitHub repository: \url{https://github.com/ds2010/pyStoNED}.
}

The purpose of this paper is to present a tutorial of the \pkg{pyStoNED} package, briefly review the alternative models supported, and illustrate its application. We focus on the estimation of frontier cost and production functions, which currently forms the main application area of these techniques, emphasizing that the various modules of the \pkg{pyStoNED} package are directly applicable for (semi-)nonparametric regression analysis in any other contexts as well. We emphasize that the \pkg{pyStoNED} package is an ongoing development by the users for the users: further model specifications and methodological advances will be implemented and added to the \pkg{pyStoNED} package on a continuous basis. 

The paper is organized as follows. Section \ref{sec:setup} describes the basic setups of the \pkg{pyStoNED} package, and Section \ref{sec:data} introduces the structures of example data and the attributes of different models. Section \ref{convexR} describes the first step of the StoNED model (e.g., the CNLS estimator) to estimate the conditional mean and some other commonly used extensions. The \proglang{Python} code for the CNLS estimator and these extensions are included. Section \ref{sec:stoned} demonstrates the rest of the steps of the StoNED model and related codes. Section \ref{sec:cnlsg} illustrates how to implement the CNLS-G algorithm to calculate the CNLS estimator using the developed \pkg{pyStoNED} package. The plot of the estimated function can be found in Section \ref{sec:plot}. Section \ref{sec:conc} concludes this paper. The list of acronyms is presented in Appendix \ref{app:acrony}. Appendices \ref{app:est_res_gams} and \ref{app:est_res_py} present the estimated residuals for a same CNLS model solved by \proglang{GAMS} and \pkg{pyStoNED}, respectively.

\section{Setup} \label{sec:setup}

\subsection{Installation}

The \pkg{pyStoNED} package supports Python 3.8 or later versions on Linux, macOS, and Windows, and is freely available on the Python Package Index (PyPI) at \url{https://pypi.org/project/pystoned}. The package can be installed by any of the following three approaches:
\begin{enumerate}[label= \arabic*), leftmargin=0.5cm]
    \item Install from the PyPI\\
            \code{pip install pystoned}
    \item Install from the GitHub repository \\
            \code{pip install -U git+https://github.com/ds2010/pyStoNED}
    \item Clone developing branches and install with \proglang{Python} using the \code{setup.py} script (for developers) \\
            \code{git clone https://github.com/ds2010/pyStoNED.git} \\
            \code{python setup.py install}
\end{enumerate}

The \pkg{pyStoNED} package is built based on a few existing dependencies.\footnote{
List of dependencies used in \pkg{pyStoNED}: \pkg{NumPy} (\citeauthor{harris2020}, \citeyear{harris2020}), \pkg{pandas} (\citeauthor{reback2020}, \citeyear{reback2020}), \pkg{SciPy} (\citeauthor{2020SciPy-NMeth}, \citeyear{2020SciPy-NMeth}), \pkg{Matplotlib} (\citeauthor{Hunter2007}, \citeyear{Hunter2007}), and \pkg{Pyomo} (\citeauthor{bynum2021pyomo}, \citeyear{bynum2021pyomo}). All the dependencies will be automatically added when installing the \pkg{pyStoNED} package. 
}
It is worth highlighting that \pkg{Pyomo} is a full-featured high-level programming language that provides a rich set of supporting libraries to program  CNLS,  StoNED, and various extensions. The other dependencies are also essential for \pkg{pyStoNED}. Specifically, \pkg{NumPy} and \pkg{pandas} are used to import input-output data, manipulate data, and export the estimation results. \pkg{SciPy} provides the algorithm to optimize the quasi-likelihood function. \pkg{Matplotlib} is then associated to plot the estimated functions.

\subsection{Solvers}

In the CNLS/StoNED framework, the existing models are either additive or multiplicative models, depending on the specification of the error term's structure. In the context of the optimization problems, the additive models are usually the QP problem with an exception of the CQR model, which is a linear programming (LP) problem, and all multiplicative models are the NLP problem. The attribute of models determines which type of solver will meet our needs, i.e., QP/LP-solver or NLP-solver. In the \pkg{pyStoNED} package, we can import these two kinds of solvers locally or remotely.

\subsubsection{Remote Optimization}

With the help of Network-Enabled Optimization System (NEOS) Server that freely provides a larger number of academic solvers such as \pkg{CPLEX} (\citeauthor{cplex2009}, \citeyear{cplex2009}), \pkg{MOSEK} (\citeauthor{mosek}, \citeyear{mosek}), \pkg{MINOS} (\citeauthor{murtagh2012}, \citeyear{murtagh2012}), and \pkg{KNITRO} (\citeauthor{byrd2006}, \citeyear{byrd2006}),\footnote{
NEOS Server: State-of-the-Art Solvers for Numerical Optimization, \url{https://neos-server.org/neos}
}
\pkg{pyStoNED} has the ability to calculate the StoNED related models without installing the local solvers.  Here we have a model optimizing remotely:
\begin{CodeChunk}
\begin{CodeInput}
>>> model.optimize(email="email@address", solver='mosek')
\end{CodeInput}
\end{CodeChunk}

The use of the remote optimization requiring \code{email} and \code{solver} as parameters.\footnote{
Replace with your own email address required by NEOS server, see \url{https://neos-guide.org/content/FAQ\#email}
}
If the user leaves the parameter \code{solver} out (i.e., \code{model.optimize(email="email@address")}), the optimization job will be calculated by the default solver. That is, the additive and multiplicative models will be solved by \pkg{MOSEK} and \pkg{KNITRO}, respectively. The remote optimization is highly recommended for light computing jobs.

\subsubsection{Local Optimization}

We can resort to the local solver pre-installed by ourselves to estimate the StoNED related models (e.g., \pkg{MOSEK} and \pkg{CPLEX}). The parameter \code{OPT_LOCAL} is added in the function \code{.optimize(...)} to indicate that the model optimizes locally. Here is an example of optimizing the additive model with local solver \pkg{MOSEK}.\footnote{
The tutorial of the \pkg{MOSEK} installation is available at \url{https://pystoned.readthedocs.io/en/latest/install/index.html}.
}
\begin{CodeChunk}
\begin{CodeInput}
>>> model.optimize(OPT_LOCAL, solver='mosek')
\end{CodeInput}
\end{CodeChunk}

Since the free and stable NLP solver remains scant, the \pkg{pyStoNED} package does not support the multiplicative model estimation locally at present. Hence, we recommend the user solve their multiplicative models remotely. We will update the package immediately when the free and stable NLP solver is available.

\section{Data structures and dataset} \label{sec:data}

\subsection{Data structures}
Data pre-processing is the first step to use the developed package, and thus the user must prepare the dataset based on the data structures of \pkg{pyStoNED}. In all CNLS/StoNED models, two common vectors or matrix are required: input variables $\bm{x}_i$ and output variables $\bm{y}_i$ for observed Decision Making Units (DMU) $i = 1, 2, \cdots, n$.  For a model considering operational conditions and practices, contextual variables $\bm{z}_i$ are required. The directional distance function (DDF) based models handle the multi-dimensional inputs and outputs with given directional vectors $\bm{g}^x$ and $\bm{g}^y$, respectively. The DDF based models also take undesired outputs $\bm{b}_i$ and its directional vector $\bm{g}^b$ into consideration. Table \ref{tab:tab1} summarizes the data structures used in all the CNLS/StoNED models.
\begin{table}[!htbp]
\begin{center}
\begin{tabular}{lllp{7.4cm}}
\hline
Symbol     & Model                   & Description \\ \hline
$\bm{x}$   & All models              & Input variables \\
$\bm{y}$   & All models              & Output variables \\
$\bm{z}$   & Contextual based models & Contextual variables \\
$\bm{b}$   & DDF based models        & Undesirable outputs \\
$\bm{g}^x$ & DDF based models        & The direction of inputs \\
$\bm{g}^y$ & DDF based models        & The direction of outputs \\
$\bm{g}^b$ & DDF based models        & The direction of undesirable outputs \\
\hline
\end{tabular} 
\caption{Data structures.}
\label{tab:tab1}
\end{center}
\vspace{-2em}
\end{table}

The inputs of the estimation functions support two forms: \code{matrix} from \pkg{NumPy} and \code{list} from \proglang{Python}. The outputs (i.e., final estimates) are retrieved by using the corresponding functions \code{.get_xxx()} provided by \pkg{pyStoNED}, all of which are in the form of \code{ndarray} from \pkg{NumPy}.

\subsection{Internal data}
To illustrate the application of the \pkg{pyStoNED} package, four commonly used datasets are attached:
\begin{enumerate}[label= \arabic*), leftmargin=0.5cm]
    \setlength{\itemsep}{1pt}
    \setlength{\parskip}{0pt}
    \setlength{\parsep}{0pt}
    \item Regulation of Finnish electricity distribution firms (\code{load_Finnish_electricity_firm}) \\
    The data of the regulation of Finnish electricity distribution firms are collected from \citet{Kuosmanen2012b} and \citet{Kuosmanen2013}. The data consist of seven variables: three different expenditures are used as inputs (i.e., \code{OPEX}, \code{CAPEX}, and \code{TOTEX});\footnote{
    Note that \code{TOTEX} = \code{OPEX} + \code{CAPEX}. It is possible to use \code{TOTEX} as an aggregate input, or model \code{OPEX} and \code{CAPEX} as two separate input variables.
    }
    \code{Energy}, \code{Length}, and \code{Customers} are considered as outputs; Further, \code{PerUndGr} is denoted as the contextual variable. Table \ref{tab:tab2} presents the description of the dataset.
    \begin{table}[!htbp]
        \begin{center}
        \begin{tabular}{lllp{7.4cm}}
        \hline
        Variable         & Unit              & Description \\ \hline
        \code{OPEX}      & Thousand Euro     & Controllable operational expenditure \\
        \code{CAPEX}     & Thousand Euro     & Total capital expenditure \\
        \code{TOTEX}     & Thousand Euro     & Total expenditure \\
        \code{Energy}    & Gigawatt Hours    & Weighted amount of energy transmitted \\
        \code{Length}    & Kilometer         & Length of the network \\
        \code{Customers} & Person            & Customers connected to the network \\
        \code{PerUndGr}  & Percentage        & Proportion of underground cabling\\
        \hline
        \end{tabular}
        \caption{The regulation of Finnish electricity distribution firms.}
        \label{tab:tab2}
        \end{center}
        \vspace{-2em}
    \end{table}
    \item GHG abatement cost of OECD countries (\code{load_GHG_abatement_cost})\\
    The data of the Greenhouse gas (GHG) abatement cost of OECD countries are provided by \citet{Kuosmanen2020}. The data contain two input variables (i.e., \code{CPNK} and \code{HRSN}), one good output variable (i.e., \code{VALK}), and one undesirable output variable (i.e., \code{GHG}). Table \ref{tab:tab3} describes the dataset in detail.
    \begin{table}[!htbp]
        \begin{center}
        \begin{tabular}{lllp{7.4cm}}
        \hline
        Variable    & Unit                               & Description \\ \hline
        \code{CPNK} & Billion Euro$^{2010}$              & Net capital stock\\
        \code{HRSN} & Billion hours                      & Hours worked by total engaged\\
        \code{VALK} & Billion Euro$^{2010}$              & Value added\\
        \code{GHG}  & Million tons of CO$_2$ equivalents & Total GHG emissions\\
        \hline
        \end{tabular}
        \caption{The GHG abatement cost of OECD countries.}
        \label{tab:tab3}
        \end{center}
        \vspace{-2em}
    \end{table}
    \item Data provided with Tim Coelli's Frontier 4.1 (\code{load_Tim_Coelli_frontier})\\
    The classic 60 firms dataset attached in \proglang{Frontier 4.1} (\citeauthor{coelli1996}, \citeyear{coelli1996}) includes two input variables (i.e., \code{capital} and \code{labour}) and one output variable (i.e., \code{output}) (see Table \ref{tab:tab4}).
    \begin{table}[h!]
        \begin{center}
        \begin{tabular}{lllp{7.4cm}}
        \hline
        Variable       & Unit           & Description \\ \hline
        \code{firm}    & Quantity       & Firm ID\\
        \code{output}  & Quantity Index & Output quantity\\
        \code{capital} & Quantity Index & Capital input\\
        \code{labour}  & Quantity Index & Labour input\\
        \hline
        \end{tabular}
        \caption{The data provided with Tim Coelli's Frontier 4.1.}
        \label{tab:tab4}
        \end{center}
        \vspace{-2em}
    \end{table}
    \item Rice Production in the Philippines (\code{load_Philipines_rice_production})\\
    The Rice Production in the Philippines dataset collected from \citet{coelli2005} consists of 17 different variables. The different variables can be organized into diversified combinations for target models. Table \ref{tab:tab5} summarizes the variables of the data.
    \begin{table}[!htbp]
        \begin{center}
        \begin{tabular}{lllp{7.5cm}}
        \hline
        Variable        	& Unit          & Description \\ \hline
        \code{YEARDUM}      & Year          & Time period \\
        \code{FMERCODE}     & Quantity      & Farmer code \\
        \code{PROD}     	& Tonnes        & Toones of freshly threshed rice\\
        \code{AREA}    		& Hactares      & Area planted\\
        \code{LABOR}    	& Mandays       & Labour used\\
        \code{NPK}      	& Kilogram      & Fertiliser used\\
        \code{OTHER}    	& Laspeyres\ index & Other inputs used\\ 
        \code{PRICE}   		& Pesos/kilogram   & Output price\\
        \code{AREAP}   		& Pesos/hectare    & Rental price of land\\
        \code{LABORP}   	& Pesos/day        & Labour price\\
        \code{NPKP}         & Pesos/kilogram   & Fertiliser price\\ 
        \code{OTHERP}       & Implicit\ price\ index & Price of other inputs\\
        \code{AGE}          & Years                  & Age of the household head\\
        \code{EDYRS}        & Years                  & Education of the household head\\
        \code{HHSIZE}       & Quantity               & Household size\\
        \code{NADULT}       & Quantity               & Number of adults in the household \\
        \code{BANRAT}       & Percentage             & Percentage of area classified as upland fields \\
        \hline
        \end{tabular}
        \caption{The rice production in the Philippines.}
        \label{tab:tab5}
        \end{center}
        \vspace{-2em}
    \end{table}
\end{enumerate}

These datasets can be imported through the module \code{dataset} of \pkg{pyStoNED}. The variables in Tables \ref{tab:tab2}--\ref{tab:tab5} are used as parameters of \code{dataset} to load the input data. The following example demonstrates how to load the input-output data from the regulation of Finnish electricity distribution firms dataset. We first import the dataset module using the function \code{load_Finnish_electricity_firm} (i.e., Line 1),\footnote{
In addition to the presented example, we can load other datasets using the corresponding dataset module, e.g., loading the GHG data: \code{from pystoned.dataset import load\_GHG\_abatement\_cost}.
}
then define the $\bm{x}$ and $y$ according to the imported dataset (Line 2). We can check the input and output data using the function \code{print(...)} (Lines 3 and 4). The parameters \code{x_select}, \code{y_select}, and \code{z_select} in \code{load_Finnish_electricity_firm(x_select, y_select, z_select)} are used to select the inputs, outputs, and  contextual variables, respectively. 
\begin{CodeChunk}
\begin{CodeInput}
>>> from pystoned.dataset import load_Finnish_electricity_firm
>>> data=load_Finnish_electricity_firm(x_select=['Energy','Length','Customers'],
...     y_select=['TOTEX'], z_select=['PerUndGr'])
>>> print(data.x)
>>> print(data.y)
>>> print(data.z)
\end{CodeInput}
\begin{CodeOutput}
[[    75    878   4933]
    .
    .
 [   105    575   9084]]
\end{CodeOutput}
\begin{CodeOutput}
[[  1612]
    .
    .
 [  1776]]
\end{CodeOutput}
\begin{CodeOutput}
[[0.11]
    .
    .
 [0.59]]
\end{CodeOutput}
\end{CodeChunk}

Note that the parameters in the module \code{dataset} can be defined according to the user's purpose. For example, if the target model only consists of two inputs (e.g., \code{Energy} and \code{Customers}) and one output (e.g., \code{TOTEX}), then the data selection in Line 2 should be 
\begin{CodeChunk}
\begin{CodeInput}
>>> data=load_Finnish_electricity_firm(x_select=['Energy','Customers'], 
...     y_select=['TOTEX'])
\end{CodeInput}
\end{CodeChunk}

\subsection{External data}
In practice, the user's own dataset is the main input of \pkg{pyStoNED}. We present an example to show how to import the user's own data. Assume that Table \ref{tab:tab6} is the input-output data stored in Excel file \code{table1.xlsx}, the following code utilizes the \pkg{pandas} to read the Excel file and organize the data with \pkg{NumPy}. The input variable \code{x} is a \code{matrix} and the output variable \code{y} is then an \code{array}.
\begin{table}[!htbp]
\begin{center}
\begin{tabular}{llll}
\hline
id  & output & input1 & input2   \\ \hline
1   & 120    & 10     & 55       \\
2   &  80    & 30     & 49       \\
.   &   .    &  .     &  .       \\
.   &   .    &  .     &  .       \\
100 &  90    & 25     & 72       \\
\hline
\end{tabular}
\caption{An example of user's own dataset.}
\label{tab:tab6}
\end{center}
\vspace{-2em}
\end{table}

In the following case, we first import the packages \pkg{NumPy} and \pkg{pandas} in Lines 1 and 2. Line 3 is used to read the Excel data file and the rest of the lines define the input and output variables. See more similar examples in Section \ref{sec:cnlsg}.
\begin{CodeChunk}
\begin{CodeInput}
>>> import numpy as np
>>> import pandas as pd
>>> df=pd.read_excel("table1.xlsx")
>>> y=df['output']
>>> x1=df['input1']
>>> x1=np.asmatrix(x1).T
>>> x2= df['input2']
>>> x2=np.asmatrix(x2).T
>>> x=np.concatenate((x1, x2), axis=1)
\end{CodeInput}
\end{CodeChunk}

\section{Shape-constrained nonparametric regression}\label{convexR}

Consider a standard multivariate, cross-sectional model in production economics:
\begin{alignat}{2}
    y_i & = f(\bm{x}_i) + \varepsilon_i  &{\quad}& \label{eq:eq1}\\
        & = f(\bm{x}_i) + v_i - u_i  &{\quad}& \forall i \notag
\end{alignat}
where $y_i$ is the output of the DMU $i$, $f: R_+^m \rightarrow R_+$ is the production (cost) function that characterizes the production (cost) technology, and $\bm{x}_i = (x_{i1}, x_{i2}, \cdots, x_{im})^{'}$ denotes the input vector of unit $i$. Similar to the literature in SFA, for production function, the presented composite error term $\varepsilon_i = v_i - u_i$ consists of the inefficiency term $u_i>0$ and stochastic noise term $v_i$. Note that by setting $u_i = 0$ we have the standard nonparametric regression model as a special case. To estimate the function $f$, one could resort to the parametric and nonparametric methods or neoclassical and frontier models, which are classified based on the specifications of $f$ and error term $\varepsilon$ (see \citeauthor{Kuosmanen2010a}, \citeyear{Kuosmanen2010a}). In this paper, we assume certain axiomatic properties (e.g., monotonicity, concavity) instead of \textit{a priori} functional form for the function $f$ and apply the following nonparametric methods to estimate the function $f$.\footnote{
In this paper, we focus on introducing the CNLS/StoNED and other related models and illustrating how to apply the \pkg{pyStoNED} to solve these models. Of course, one can use this package to solve other nonparametric models such as DEA and FDH. See more tutorials about DEA and FDH estimations at \url{https://pystoned.readthedocs.io}.
}

\subsection{Convex Nonparametric Least Square} \label{sec:CNLS}
\subsubsection{Additive CNLS model}

\citet{Hildreth1954} is the first to consider the nonparametric regression subject to monotonicity and concavity constraints in the case of a single input variable $x$. \citet{Afriat1972} also proposes methods to impose convexity on the estimation of a production function. \citet{Kuosmanen2008} extends Hildreth's approach to the multivariate setting with the multidimensional input $\bm{x}$, and refers it to as the CNLS. CNLS builds upon the assumption that the true but unknown production function $f$ belongs to the set of continuous, monotonic increasing and globally concave (convex) functions, imposing the same production axioms as standard DEA (see further discussion in \citeauthor{Kuosmanen2010a}, \citeyear{Kuosmanen2010a}). The additive multivariate CNLS formulations are defined as
\begin{itemize}
    \item Estimating production function (i.e., regression function $f$ is concave and increasing)
        \begin{alignat}{2}
            \underset{\alpha, \bm{\beta}, \varepsilon} \min &\sum_{i=1}^n\varepsilon_i^2 &{\quad}& \label{eq:eq2}\\
            \mbox{\textit{s.t.}}\quad 
            &  y_i = \alpha_i + \bm{\beta}_i^{'}\bm{x}_i + \varepsilon_i &{\quad}& \forall i \notag \\
            &  \alpha_i + \bm{\beta}_i^{'}\bm{x}_i \le \alpha_j + \bm{\beta}_j^{'}\bm{x}_i  &{\quad}&  \forall i, j ,\; \text{and} \; i \neq j\notag\\
            &  \bm{\beta}_i \ge \bm{0} &{\quad}& \forall i \notag
            \end{alignat}
    \item Estimating cost function (i.e., regression function $f$ is convex and increasing)
        \begin{alignat}{2}
            \underset{\alpha, \bm{\beta}, \varepsilon} \min & \sum_{i=1}^n\varepsilon_i^2 &{\quad}& \label{eq:eq3}\\
            \mbox{\textit{s.t.}}\quad 
            &  y_i = \alpha_i + \bm{\beta}_i^{'}\bm{x}_i + \varepsilon_i &{\quad}& \forall i  \notag \\
            &  \alpha_i + \bm{\beta}_i^{'}\bm{x}_i \ge \alpha_j + \bm{\beta}_j^{'}\bm{x}_i  &{\quad}&  \forall i, j ,\; \text{and} \; i \neq j\notag\\
            &  \bm{\beta}_i \ge \bm{0} &{\quad}& \forall i \notag
        \end{alignat}
\end{itemize}
where $\alpha_i$ and $\bm{\beta}_i$ define the intercept and slope parameters of tangent hyperplanes that characterize the estimated piecewise linear frontier, respectively. $\varepsilon_i$ denotes the regression residual. The first constraint can be interpreted as a multivariate regression equation, the second constraint imposes convexity (concavity), and the third constraint imposes monotonicity. Similar to the DEA specification, other standard specifications of returns to scale can be imposed by an additional constraint on the intercept term $\alpha_i$. If $\alpha_i=0$, then Problems \eqref{eq:eq2} and \eqref{eq:eq3} are the constant returns to scale (CRS) model, otherwise they are the variable returns to scale (VRS) model. Note that both Problems \eqref{eq:eq2} and \eqref{eq:eq3} are the QP problem and hence can be solved by \pkg{MOSEK} and \pkg{CPLEX}.

The basic additive CNLS model can be estimated in \pkg{pyStoNED} using the module \code{CNLS(y, x, ...)} with the contextual variable \code{z} parameter set to \code{None} (default) and the type of model \code{cet} parameter set to \code{CET_ADDI} (additive model; default). The type of estimated function can be classified by setting the \code{fun} parameter to \code{FUN_PROD} (production function; default) or \code{FUN_COST} (cost function). The returns to scale assumption can be specified by setting the \code{rts} parameter to \code{RTS_VRS} (VRS model; default) or \code{RTS_CRS} (CRS model). The results can be displayed in the screen directly using the \code{.display_alpha()} (i.e., display the coefficients $\hat{\alpha}_i$) or stored in the memory using the \code{.get_alpha()}.

We first present an example to solve the VRS production model and store the estimates as the following Lines 12-14.
\begin{CodeChunk}
\begin{CodeInput}
>>> from pystoned import CNLS
>>> from pystoned.constant import CET_ADDI, FUN_PROD, OPT_LOCAL, RTS_VRS
>>> from pystoned.dataset import load_Finnish_electricity_firm
>>> data = load_Finnish_electricity_firm(x_select=['OPEX','CAPEX'],
...     y_select=['Energy'])
>>> model=CNLS.CNLS(y=data.y, x=data.x, z=None, 
...     cet=CET_ADDI, fun=FUN_PROD, rts=RTS_VRS)
>>> model.optimize(OPT_LOCAL)
>>> 
>>> model.display_alpha()
>>> model.display_beta()
>>> model.display_residual()
>>> 
>>> alpha = model.get_alpha()
>>> beta = model.get_beta()
>>> residuals = model.get_residual()
\end{CodeInput}
\begin{CodeOutput}
alpha : alpha
    Size=89, Index=I
    Key : Lower : Value               : Upper : Fixed : Stale : Domain
      0 :  None : -22.935939927355427 :  None : False : False :  Reals
      1 :  None :  -22.98683384275405 :  None : False : False :  Reals
      .
      .
     88 :  None : -22.860373724764905 :  None : False : False :  Reals
\end{CodeOutput}
\begin{CodeOutput}
beta : beta
    Size=178, Index=beta_index
    Key     : Lower : Value                  : Upper : Fixed : Stale : Domain
     (0, 0) :   0.0 :    0.13585845731547977 :  None : False : False :  Reals
     (0, 1) :   0.0 :   0.011273984231181894 :  None : False : False :  Reals
     (1, 0) :   0.0 :    0.13637662344803758 :  None : False : False :  Reals
     (1, 1) :   0.0 :   0.010903786808952633 :  None : False : False :  Reals
        .
        .
    (88, 0) :   0.0 :    0.13561549262636433 :  None : False : False :  Reals
    (88, 1) :   0.0 :   0.011417220278201383 :  None : False : False :  Reals
\end{CodeOutput}
\begin{CodeOutput}
epsilon : residual
    Size=89, Index=I
    Key : Lower : Value               : Upper : Fixed : Stale : Domain
      0 :  None : -2.8024040090178914 :  None : False : False :  Reals
      1 :  None :  1.4140528128759229 :  None : False : False :  Reals
      .
      .
     88 :  None :  -0.885163858119256 :  None : False : False :  Reals
\end{CodeOutput}
\end{CodeChunk}
In this example, Lines 1-3 import the CNLS module, parameter setting modules, and dataset module. Line 4 defines the input and output variables using the Finnish electricity distribution firm data. Lines 5-6 define the CNLS production model with a user-defined name (e.g., \code{model}) and solve the production model using the local off-the-shelf solver (i.e., \pkg{MOSEK}).\footnote{
The name of estimated model is free to be assigned, e.g., \code{CNLS\_model=CNLS.CNLS(y=data.y, x=data.x, z=None, cet=CET\_ADDI, fun=FUN\_PROD, rts=RTS\_VRS)}. Correspondingly, the Line 6 in this example should be replaced by \code{CNLS\_model.optimize(OPT\_LOCAL)}, and Line 8 now is \code{CNLS\_model.display\_alpha()}. \label{fn:fn1}
}
Lines 8-10 directly display the estimated coefficients (i.e., $\hat{\alpha}_i$, $\hat{\beta}_{ij}$, and $\hat{\varepsilon}_i$) on screen, and Lines 12-14 store the estimates in memory with a special variable name (e.g, alpha). Appendix \ref{app:est_res_py} presents the full estimated CNLS residuals that are equivalent to those in \proglang{GAMS} (cf. Appendix \ref{app:est_res_gams}). Note that the estimated alpha and beta in \pkg{pyStoNED} may be slightly different from those in \proglang{GAMS} due to non-uniqueness in CNLS estimation (see \citeauthor{Kuosmanen2008}, \citeyear{Kuosmanen2008}; \citeauthor{Dai2021}, \citeyear{Dai2021}).

\subsubsection{Multiplicative CNLS model}

Similar to the existing SFA literature, the Cobb-Douglas and translog functions are commonly assumed to be the functional form for the function $f$, where inefficiency $u$ and noise $v$ affect production in a multiplicative fashion. We, thus, further consider the multiplicative specification in the nonparametric models. Note that the assumption of CRS would also require a multiplicative error structure. Under the multiplicative composite error structure, the function \eqref{eq:eq1} is rephrased as
\begin{align}
    y_i = f(\bm{x}_i) \cdot \exp{(\varepsilon_i)} = f(\bm{x}_i) \cdot \exp{( v_i - u_i)}  \label{eq:eq4}
\end{align}
Applying the log-transformation to Eq.\eqref{eq:eq4}, we obtain
\begin{align}
    \ln y_i = \ln f(\bm{x}_i) + v_i - u_i \label{eq:eq5}
\end{align}
To estimate Eq.\eqref{eq:eq5}, we reformulate the additive production model \eqref{eq:eq2} and obtain the following log-transformed CNLS formulation:
\begin{alignat}{2}
    \underset{\alpha, \bm{\beta}, \phi, \varepsilon} \min & \sum_{i=1}^n\varepsilon_i^2  &{\quad}& \label{eq:eq6}\\
    \mbox{\textit{s.t.}}\quad 
    &  \ln y_i = \ln(\phi_i+1) + \varepsilon_i  &{\quad}& \forall i  \notag\\
    & \phi_i  = \alpha_i+\bm{\beta}_i^{'}\bm{x}_i -1 &{\quad}& \forall i  \notag \\
    &  \alpha_i + \bm{\beta}_i^{'}\bm{x}_i \le \alpha_j + \bm{\beta}_j^{'}\bm{x}_i  &{\quad}&  \forall i, j ,\; \text{and} \; i \neq j\notag\\
    &  \bm{\beta}_i \ge \bm{0} &{\quad}&  \forall i  \notag 
\end{alignat}
where $\phi_i+1$ is the CNLS estimator of $\E[y_i \, | \, \bm{x}_i]$. The value of one is added here to make sure that the computational algorithms do not take the logarithm of zero. The first equality can be interpreted as the log-transformed regression equation (using the natural logarithm function $\ln(\cdot)$). The rest of the constraints are the same as those of the additive models. The use of $\phi_i$ allows the estimation of a multiplicative relationship between output and input while assuring convexity of the production possibility set in the original input-output space. Note that one could not apply the log transformation directly to the input data $\bm{x}$ due to the fact that the piece-wise log-linear frontier does not satisfy the axiomatic property (i.e., concavity or convexity) of the function $f$. Since the multiplicative model \eqref{eq:eq6} is a nonlinear programming (NLP) problem, we need to use the nonlinear solvers, e.g., \pkg{MINOS}, \pkg{KNITRO}. 

We next demonstrate the estimation of both VRS and CRS multiplicative cost function models. Let the type of model \code{cet} parameter be \code{CET\_MULT} (multiplicative model). Note that the following NLP models are remotely solved by \pkg{KNITRO} solver via the NEOS Server. 
\begin{CodeChunk}
\begin{CodeInput}
>>> from pystoned import CNLS
>>> from pystoned.constant import CET_MULT, FUN_COST, RTS_VRS, RTS_CRS
>>> from pystoned.dataset import load_Finnish_electricity_firm
>>> data = load_Finnish_electricity_firm(x_select=['Energy','Length','Customers'],
...     y_select=['TOTEX'])
>>> model1=CNLS.CNLS(y=data.y, x=data.x, z=None, 
...     cet=CET_MULT, fun=FUN_COST, rts=RTS_VRS)
>>> model1.optimize('email@address')
>>> model2=CNLS.CNLS(y=data.y, x=data.x, z=None, 
...     cet=CET_MULT, fun=FUN_COST, rts=RTS_CRS)
>>> model2.optimize('email@address')
>>> 
>>> model1.display_residual()
>>> model2.display_residual()
\end{CodeInput}
\begin{CodeOutput}
epsilon : residual
    Size=89, Index=I
    Key : Lower : Value                 : Upper : Fixed : Stale : Domain
      0 :  None :   0.03795367428994332 :  None : False : False :  Reals
      1 :  None :  0.030099796252526297 :  None : False : False :  Reals
      .
      .
     88 :  None :   0.04860152062781719 :  None : False : False :  Reals
\end{CodeOutput}
\begin{CodeOutput}
epsilon : residual
    Size=89, Index=I
    Key : Lower : Value                  : Upper : Fixed : Stale : Domain
      0 :  None :    0.03591704717267953 :  None : False : False :  Reals
      1 :  None :   0.026035484542682643 :  None : False : False :  Reals
      .
      .
     88 :  None :    0.07738478945809676 :  None : False : False :  Reals
\end{CodeOutput}
\end{CodeChunk}

\subsubsection{Corrected CNLS model}

Corrected convex nonparametric least squares (C$^2$NLS) is a variant of the corrected ordinary least squares (COLS) model, in which nonparametric least squares subject to monotonicity and concavity constraints replace the first-stage parametric ordinary least squares (OLS) regression. To estimate the production function, the C$^2$NLS model assumes that the regression $f$ is monotonic increasing and globally concave production function, the inefficiencies $\varepsilon$ are identically and independently distributed with mean $\mu$ and a finite variance $\sigma^2$, and that the inefficiencies $\varepsilon$ are uncorrelated with inputs $\bm{x}$.

Similar to COLS, the C$^2$NLS method includes two stages, which can be stated as follows:
\begin{itemize}
    \item Estimate $\E[y_i \, | \, x_i] $ by solving the CNLS model, e.g., Problem \eqref{eq:eq2}. Denote the CNLS residuals by $\varepsilon^{CNLS}_i$.
    \item Shift the residuals analogous to the COLS procedure; the C$^2$NLS efficiency estimator is
    $$\hat{\varepsilon}_i^{C2NLS}= \varepsilon_i^{CNLS} - \max_j \varepsilon_j^{CNLS} $$
    where values of $\hat{\varepsilon_i}^{C2NLS}$ range from $[0, +\infty]$ with 0 indicating efficient performance. Similarly, we adjust the CNLS intercepts $\alpha_i$ as 
    $$ \hat{\alpha}_i^{C2NLS}= \alpha_i^{CNLS} + \max_j \varepsilon_j^{CNLS} $$
    where $\alpha^{CNLS}_i$ is the optimal intercept for firm $i$ in above CNLS problem and $\hat{\alpha}_i^{C2NLS}$ is the C$^2$NLS estimator. Slope coefficients $\bm{\beta}_i$ for C$^2$NLS are obtained directly as the optimal solution to the CNLS problem.
\end{itemize}
To solve the C$^2$NLS model (e.g., additive production model), we only need to introduce two new functions: \code{.get_adjusted_residual()} and \code{.get_adjusted_alpha()}. After completing the first-stage estimation, we use these two functions to obtain the adjusted residuals and intercept terms. 
\begin{CodeChunk}
\begin{CodeInput}
>>> from pystoned import CNLS
>>> from pystoned.constant import CET_ADDI, FUN_PROD, OPT_LOCAL, RTS_VRS
>>> from pystoned.dataset import load_Finnish_electricity_firm
>>> data = load_Finnish_electricity_firm(x_select=['OPEX', 'CAPEX'], 
...     y_select=['Energy'])
>>> model=CNLS.CNLS(y=data.y, x=data.x, z=None, 
...     cet=CET_ADDI, fun=FUN_PROD, rts=RTS_VRS)
>>> model.optimize(OPT_LOCAL)
>>> 
>>> print(model.get_adjusted_residual())  
>>> print(model.get_adjusted_alpha())  
\end{CodeInput}
\begin{CodeOutput}
[ -682.18627782  -677.969821    -701.60761871 -1030.2948489
  -693.08377375  ...            -685.44491325 -680.26903767]
\end{CodeOutput}
\begin{CodeOutput}
[ 656.44793389  656.39703997  656.51998647  712.844865   
  656.44540447  ...           656.38925598  656.52350009]
\end{CodeOutput}
\end{CodeChunk}

\subsection{Convex Quantile and Expectile Regression} \label{sec:CQR}

While CNLS estimates the conditional mean $\E(y_i \, | \, x_i)$, quantile regression aims at estimating the conditional median or other quantiles of the response variable (\citeauthor{Koenker1978}, \citeyear{Koenker1978}; \citeauthor{Koenker2005b}, \citeyear{Koenker2005b}) and provides an overall picture of the conditional distributions. The quantile $\tau$ splits the observations $\tau$\% above and (1-$\tau$)\% below. In this subsection, we extend CNLS problem to estimate CQR (\citeauthor{Wang2014c}, \citeyear{Wang2014c}; \citeauthor{Kuosmanen2015d}, \citeyear{Kuosmanen2015d}) and convex expectile regression (CER) (\citeauthor{Kuosmanen2020}, \citeyear{Kuosmanen2020}; \citeauthor{Dai2020}, \citeyear{Dai2020}; \citeauthor{Kuosmanen2021}, \citeyear{Kuosmanen2021}). Note that both quantile and expectile estimations are more robust to outliers and heteroscedasticity than the CNLS estimation.

\subsubsection{Convex quantile regression}

Given a pre-specified quantile $\tau \in (0, 1)$, the CQR model is formulated as 
\begin{alignat}{2}
    \underset{\alpha,\bm{\beta},{{\varepsilon}^{\text{+}}},{{\varepsilon}^{-}}}{\mathop{\min }}&\,
    \tau \sum\limits_{i=1}^{n}{\varepsilon _{i}^{+}}+(1-\tau )\sum\limits_{i=1}^{n}{\varepsilon _{i}^{-}}  &{}&  \label{eq:eq7}\\ 
    \mbox{\textit{s.t.}}\quad 
    & y_i=\alpha_i+ \bm{\beta}_i^{'}\bm{x}_i+\varepsilon _i^{+} - \varepsilon _i^{-} &{}& \forall i \notag \\
    & \alpha_i+\bm{\beta}_i^{'}\bm{x}_i \le \alpha_j+\bm{\beta}_j^{'}\bm{x}_i  &{}& \forall i, j ,\; \text{and} \; i \neq j\notag\\
    & \bm{\beta}_i\ge \bm{0} &{}& \forall i \notag  \\
    & \varepsilon _i^{+}\ge 0,\ \varepsilon_i^{-} \ge 0 &{}& \forall i \notag 
\end{alignat}
where $\varepsilon^{+}_i$ and $\varepsilon^{-}_i$ denote the two non-negative components. The objective function in \eqref{eq:eq7} minimizes the asymmetric absolute deviations from the function rather than the symmetric quadratic deviations. The last set of constraints is the sign constraint of the error terms. The other constraints are the same as those of the CNLS problem \eqref{eq:eq2}.

\subsubsection{Convex expectile regression}

Convex quantile regression \eqref{eq:eq7} may suffer from non-uniqueness due to that Problem \eqref{eq:eq7} is a linear programming (LP) problem (\citeauthor{Kuosmanen2015d}, \citeyear{Kuosmanen2015d}). To address this problem, \citet{Kuosmanen2015d} purpose a CER approach, where a quadratic objective function is used to ensure unique estimates of the quantile functions. Consider the following QP problem
\begin{alignat}{2}
    \underset{\alpha,\bm{\beta},{{\varepsilon}^{\text{+}}},{\varepsilon}^{-}}{\mathop{\min}}&\,
    \tilde{\tau} \sum\limits_{i=1}^n(\varepsilon _i^{+})^2+(1-\tilde{\tau} )\sum\limits_{i=1}^n(\varepsilon_i^{-})^2   &{}&  \label{eq:eq8}\\ 
    \mbox{\textit{s.t.}}\quad 
    & y_i=\alpha_i+ \bm{\beta}_i^{'}\bm{x}_i+\varepsilon _i^{+}-\varepsilon _i^{-}  &{}&  \forall i \notag  \\
    & \alpha_i+\bm{\beta}_i^{'}\bm{x}_i \le \alpha_j+\bm{\beta}_j^{'}\bm{x}_i  &{}& \forall i, j ,\; \text{and} \; i \neq j\notag\\
    & \bm{\beta}_i\ge \bm{0}  &{}&  \forall i  \notag  \\
    & \varepsilon _i^{+}\ge 0,\ \varepsilon_i^{-} \ge 0  &{}& \forall i \notag 
\end{alignat}
where the expectile $\tilde{\tau} \in (0, 1)$ can be converted from/to the quantile $\tau$ (\citeauthor{Kuosmanen2021}, \citeyear{Kuosmanen2021}). 

The alternative module \code{CQER} includes the function \code{CQR(y, x, ...)} that is designed for solving CQR problem and the function \code{CER(y, x, ...)} that is for CER problem. Therefore, we use the \code{CQER.CQR()} and \code{CQER.CER()} to define the CQR problem (see Line 5 in following example) and the CER problem, respectively. The other parameters settings are similar to those in module \code{CNLS(y, x, ...)}. To display the estimated $\varepsilon_i^{+}$ and $\varepsilon_i^{-}$, the functions \code{.display_positive_residual()} and \code{.display_negative_residual()} are designed in the new module \code{CQER}.\footnote
{Note that as shown in Problem \ref{eq:eq8}, both residuals $\varepsilon _i^{+}$ and $\varepsilon_i^{-}$ are larger than 0. The function names, positive and negative residuals, are simply used to keep the consistency with the math notations in Problem \ref{eq:eq8}.
}
The following additive CQR model is presented to estimate a quantile production function.
\begin{CodeChunk}
\begin{CodeInput}
>>> from pystoned import CQER
>>> from pystoned.constant import CET_ADDI, FUN_PROD, OPT_LOCAL, RTS_VRS
>>> from pystoned import dataset as dataset
>>> data=dataset.load_GHG_abatement_cost(x_select=['HRSN','CPNK','GHG'],
...     y_select=['VALK'])
>>> model=CQER.CQR(y=data.y, x=data.x, tau=0.5, z=None, 
...     cet=CET_ADDI, fun=FUN_PROD, rts=RTS_VRS)
>>> model.optimize(OPT_LOCAL)
>>> 
>>> model.display_alpha()
>>> model.display_beta() 
>>> model.display_positive_residual()
>>> model.display_negative_residual() 
\end{CodeInput}
\begin{CodeOutput}
alpha : alpha
    Size=168, Index=I
    Key : Lower : Value               : Upper : Fixed : Stale : Domain
      0 :  None : -1306.0485023191532 :  None : False : False :  Reals
      1 :  None : -1306.0485023191268 :  None : False : False :  Reals
      .
      .
    167 :  None :   170680.2112574062 :  None : False : False :  Reals
\end{CodeOutput}
\begin{CodeOutput}
beta : beta
    Size=504, Index=beta_index
    Key      : Lower : Value                   : Upper : Fixed : Stale : Domain
      (0, 0) :   0.0 :      21.043360329998244 :  None : False : False :  Reals
      (0, 1) :   0.0 :      0.1758315887019594 :  None : False : False :  Reals
      (0, 2) :   0.0 :                     0.0 :  None : False : False :  Reals
        .
        .
    (167, 0) :   0.0 :       13.22794884204182 :  None : False : False :  Reals
    (167, 1) :   0.0 :     0.11023334987523721 :  None : False : False :  Reals
    (167, 2) :   0.0 :       645.5693304144419 :  None : False : False :  Reals
\end{CodeOutput}
\begin{CodeOutput}
epsilon_plus : positive error term
    Size=168, Index=I
    Key : Lower : Value              : Upper : Fixed : Stale : Domain
      0 :   0.0 :                0.0 :  None : False : False :  Reals
      1 :   0.0 :                0.0 :  None : False : False :  Reals
      .
      .
    167 :   0.0 :  98359.68518583104 :  None : False : False :  Reals
\end{CodeOutput}
\begin{CodeOutput}
epsilon_minus : negative error term
    Size=168, Index=I
    Key : Lower : Value              : Upper : Fixed : Stale : Domain
      0 :   0.0 :  16157.82681402145 :  None : False : False :  Reals
      1 :   0.0 :  7158.090538541088 :  None : False : False :  Reals
      .
      .
    167 :   0.0 :                0.0 :  None : False : False :  Reals
\end{CodeOutput}
\end{CodeChunk}

\subsection{Contextual Variables}\label{sec:cv}

A firm's ability to operate efficiently often depends on operational conditions and practices, such as the production environment and the firm-specific characteristics (i.e., technology selection and managerial practices). \citet{Johnson2011, Johnson2012a} refer to both variables that characterize operational conditions and practices as contextual variables.
\begin{itemize}
    \item Contextual variables are often (but not always) \textit{external factors} that are beyond the control of firms
    \begin{itemize}
        \item Examples: competition, regulation, weather, location
        \item Need to adjust efficiency estimates for the operating environment
        \item Policymakers may influence the operating environment
    \end{itemize}
    \item Contextual variables can also be \textit{internal factors}
    \begin{itemize}
        \item Examples: management practices, ownership
        \item Better understanding of the impacts of internal factors can help the firm to improve performance
    \end{itemize}
\end{itemize}

By introducing the contextual variables $\bm{z}_i$, the multiplicative model \eqref{eq:eq5} is reformulated as a partial log-linear model to take the operational conditions and practices into account.
\begin{align}
    \ln y_i = \ln f(\bm{x_i}) + \bm{\lambda}^{'}\bm{z}_i + v_i - u_i
\end{align}
where parameter vector $\bm{\lambda}=(\lambda_1, \cdots, \lambda_r)$ represents the marginal effects of contextual variables on output. All other variables maintain their previous definitions. Similarly, we can also introduce the contextual variables to the additive model. In this subsection, we consider the multiplicative production model as our starting point.

\subsubsection{CNLS with z variables}
Following \citet{Johnson2011}, we incorporate the contextual variables in the multiplicative CNLS model and redefine it as follows:
\begin{alignat}{2}
    & \underset{\alpha, \bm{\beta}, \bm{\lambda}, \varepsilon} {\min} \sum_{i=1}^n\varepsilon_i^2  &{}&  \label{eq:eq10}\\
    \mbox{\textit{s.t.}}\quad 
    &  \ln y_i = \ln(\phi_i+1) + \bm{\lambda}^{'}\bm{z}_i + \varepsilon_i  &{}&  \forall i \notag\\
    &  \phi_i  = \alpha_i + \bm{\beta}_i^{'}\bm{x}_i -1 &{}&  \forall i \notag\\
    &  \alpha_i + \bm{\beta}_i^{'}\bm{x}_i \le \alpha_j + \bm{\beta}_j^{'}\bm{x}_i  &{}&   \forall i, j ,\; \text{and} \; i \neq j\notag\\
    &  \bm{\beta}_i \ge \bm{0} &{}&   \forall i \notag
\end{alignat}
Denote by $\hat{\bm{\lambda}}$ the coefficients of the contextual variables obtained as the optimal solution to the above nonlinear problem.  \citet{Johnson2011} examine the statistical properties of this estimator in detail, showing its unbiasedness, consistency, and asymptotic efficiency. 

The contextual variables $\bm{z}$ have been integrated into the modules \code{CNLS()} and \code{CQER()}. Further, the function \code{.display_lamda()} is used to display the marginal effect of contextual variable. In the following example, we estimate a log-transformed cost function model with z-variable. Note that we assume that the firms are constant returns to scale in this example. 
\begin{CodeChunk}
\begin{CodeInput}
>>> from pystoned import CNLS
>>> from pystoned.constant import CET_MULT, FUN_COST, RTS_CRS
>>> from pystoned.dataset import load_Finnish_electricity_firm
>>> data=load_Finnish_electricity_firm(x_select=['Energy','Length','Customers'],
...     y_select=['TOTEX'], z_select=['PerUndGr'])
>>> model=CNLS.CNLS(y=data.y, x=data.x, z=data.z, 
...     cet=CET_MULT, fun=FUN_COST, rts=RTS_CRS)
>>> model.optimize('email@address')
>>> 
>>> model.display_lamda()
\end{CodeInput}
\begin{CodeOutput}
lamda : z coefficient
    Size=1, Index=K
    Key : Lower : Value              : Upper : Fixed : Stale : Domain
      0 :  None : 0.3609227173656651 :  None : False : False :  Reals
\end{CodeOutput}
\end{CodeChunk}

\subsubsection{CER with z variables}

Following \citet{Kuosmanen2021a}, we also incorporate the contextual variable in the multiplicative CER estimation. The reformulation of CER model is 
\begin{alignat}{2}
    \underset{\alpha,\bm{\beta},\bm{\lambda}, \varepsilon^{+},\varepsilon^{-}}{\mathop{\min}}&\,
        \tilde{\tau} \sum\limits_{i=1}^n(\varepsilon _i^{+})^2+(1-\tilde{\tau} )\sum\limits_{i=1}^n(\varepsilon_i^{-})^2   &{}&  \label{eq:eq11}\\ 
    \mbox{\textit{s.t.}}\quad 
    &  \ln y_i = \ln(\phi_i+1) + \bm{\lambda}^{'}\bm{z}_i + \varepsilon _i^{+}-\varepsilon _i^{-}  &{}&  \forall i \notag\\
    &  \phi_i  = \alpha_i + \bm{\beta}_i^{'}\bm{x}_i -1 &{}&  \forall i \notag\\
    &  \alpha_i + \bm{\beta}_i^{'}\bm{x}_i \le \alpha_j + \bm{\beta}_j^{'}\bm{x}_i  &{}&   \forall i, j ,\; \text{and} \; i \neq j\notag\\
    &  \bm{\beta}_i \ge \bm{0} &{}&   \forall i \notag \\
    &  \varepsilon _i^{+}\ge 0,\ \varepsilon_i^{-} \ge 0  &{}& \forall i \notag 
\end{alignat}
The following code is prepared to solve the CER model with z-variable. We now use the function \code{CQER.CER()} to model the CER with z-variable problem and assume that the expectile $\tilde{\tau}$ is equal to 0.5. In this example, we also estimate a CRS cost function.
\begin{CodeChunk}
\begin{CodeInput}
>>> from pystoned import CQER
>>> from pystoned.constant import CET_MULT, FUN_COST, RTS_CRS
>>> from pystoned.dataset import load_Finnish_electricity_firm
>>> data=load_Finnish_electricity_firm(x_select=['Energy','Length','Customers'],
...     y_select=['TOTEX'], z_select=['PerUndGr'])
>>> model=CQER.CER(y=data.y, x=data.x, z=data.z, tau=0.5, 
...     cet=CET_MULT, fun=FUN_COST, rts=RTS_CRS)
>>> model.optimize('email@address')
>>> 
>>> model.display_lamda()
\end{CodeInput}
\begin{CodeOutput}
lamda : z coefficient
    Size=1, Index=K
    Key : Lower : Value              : Upper : Fixed : Stale : Domain
      0 :  None : 0.3609226375022837 :  None : False : False :  Reals
\end{CodeOutput}
\end{CodeChunk}

\subsection{Multiple Outputs (DDF Formulation)}

\subsubsection{CNLS with multiple outputs}

Until now, the convex regression approaches have been presented within the single output, multiple input framework. In this subsection, we describe the CNLS/CQR/CER approaches with the DDF to handle multiple-input multiple-output data (\citeauthor{Chambers1996}, \citeyear{Chambers1996}, \citeyear{Chambers1998b}). 

Consider the following QP problem (\citeauthor{Kuosmanen2017a}, \citeyear{Kuosmanen2017a})
\begin{alignat}{2}
    \underset{\alpha, \bm{\beta}, \bm{\gamma}, \varepsilon}{\mathop{\min}}&\sum_{i=1}^n\varepsilon_i^2  &{\quad}& \label{eq:eq12} \\
    \mbox{\textit{s.t.}}\quad 
    &  \bm{\gamma}_i^{'}\bm{y}_i = \alpha_i + \bm{\beta}_i^{'}\bm{x}_i - \varepsilon_i &{\quad}& \forall i \notag \\
    &  \alpha_i + \bm{\beta}_i^{'}\bm{x}_i -\bm{\gamma}_i^{'}\bm{y}_i \le \alpha_j + \bm{\beta}_j^{'}\bm{x}_i -\bm{\gamma}_j^{'}\bm{y}_i  &{\quad}&  \forall i, j ,\; \text{and} \; i \neq j\notag\\
    &  \bm{\gamma}_i^{'} g^{y}  + \bm{\beta}_i^{'} g^{x}  = 1  &{\quad}& \forall i  \notag \\ 
    &  \bm{\beta}_i \ge \bm{0}, \bm{\gamma}_i \ge \bm{0} &{\quad}& \forall i \notag
\end{alignat}
where the residual $\varepsilon_i$ represents the estimated value of $d$ ($\vec{D}(x_i,y_i,g^x,g^y)+u_i$). In addition to the same notations as the CNLS estimator, we also introduce firm-specific coefficients $\bm{\gamma_i}$ that represent marginal effects of outputs to the DDF.

The first constraint defines the distance to the frontier as a linear function of inputs and outputs. The linear approximation of the frontier is based on the tangent hyperplanes, analogous to the original CNLS formulation. The second set of constraints is the system of Afriat inequalities that impose global concavity. The third constraint is a normalization constraint that ensures the translation property. The last two constraints impose monotonicity in all inputs and outputs. It is straightforward to show that the CNLS estimator of function $d$ satisfies the axioms of free disposability, convexity, and translation property.

To perform the DDF related models, the \pkg{pyStoNED} includes the modules \code{CNLSDDF(y, x, ...)} and \code{CQERDDF(y, x, ...)}, and takes the undesirable outputs into account. We first present the CNLS-DDF models in the following two examples and then describe the CQR/CER-DDF models in next subsection. To apply the module \code{CNLSDDF(y, x, ...)}, we have to pre-define the directional vector for the parameters \code{gx}, \code{gb} (None; default)  and \code{gy} (i.e., Line 5: \code{gx=[1.0, 0.0], gb=None, gy=[0.0, 0.0, 0.0]}). The module also reports the estimates using \code{.display_alpha()}, \code{.display_beta()}, \code{.display_gamma()}, and \code{.display_residual()}.
\begin{CodeChunk}
\begin{CodeInput}
>>> from pystoned import CNLSDDF
>>> from pystoned.constant import FUN_PROD, OPT_LOCAL
>>> from pystoned.dataset import load_Finnish_electricity_firm
>>> data=load_Finnish_electricity_firm(x_select=['OPEX','CAPEX'],
...     y_select=['Energy','Length','Customers'])
>>> model=CNLSDDF.CNLSDDF(y=data.y, x=data.x, b=None, fun=FUN_PROD, 
...     gx=[1.0, 0.0], gb=None, gy=[0.0, 0.0, 0.0])
>>> model.optimize(OPT_LOCAL)
>>> 
>>> model.display_alpha()
>>> model.display_beta()
>>> model.display_gamma()
>>> model.display_residual()
\end{CodeInput}
\begin{CodeOutput}
alpha : alpha
    Size=89, Index=I
    Key : Lower : Value               : Upper : Fixed : Stale : Domain
      0 :  None : -231.06893282325376 :  None : False : False :  Reals
      1 :  None :  -212.5240841137753 :  None : False : False :  Reals
      .
      .
     88 :  None :  -330.8605696176572 :  None : False : False :  Reals
\end{CodeOutput}
\begin{CodeOutput}
beta : beta
    Size=178, Index=beta_index
    Key     : Lower : Value                  : Upper : Fixed : Stale : Domain
     (0, 0) :   0.0 :                    1.0 :  None : False : False :  Reals
     (0, 1) :   0.0 : 0.00027176268397506657 :  None : False : False :  Reals
        .
        .
    (88, 0) :   0.0 :                    1.0 :  None : False : False :  Reals
    (88, 1) :   0.0 :     0.0526234540690525 :  None : False : False :  Reals
\end{CodeOutput}
\begin{CodeOutput}
gamma : gamma
    Size=267, Index=gamma_index
    Key     : Lower : Value                  : Upper : Fixed : Stale : Domain
     (0, 0) :   0.0 :      3.806267544117393 :  None : False : False :  Reals
     (0, 1) :   0.0 :     0.2311214613059167 :  None : False : False :  Reals
     (0, 2) :   0.0 :   0.004895753020650473 :  None : False : False :  Reals
        .
        .
    (88, 0) :   0.0 :   0.015392372124732883 :  None : False : False :  Reals
    (88, 1) :   0.0 :  0.0029598467042669355 :  None : False : False :  Reals
    (88, 2) :   0.0 :    0.06644931421296872 :  None : False : False :  Reals
\end{CodeOutput}
\begin{CodeOutput}
epsilon : residuals
    Size=89, Index=I
    Key : Lower : Value               : Upper : Fixed : Stale : Domain
      0 :  None :  -62.41627631290396 :  None : False : False :  Reals
      1 :  None : -176.54976937092056 :  None : False : False :  Reals
      .
      .
     88 :  None : -10.336197186462073 :  None : False : False :  Reals
\end{CodeOutput}
\end{CodeChunk}

When considering undesirable outputs, the CNLS-DDF problem \eqref{eq:eq12} can be reformulated as
\begin{alignat}{2}
    \underset{\alpha, \bm{\beta}, \bm{\gamma}, \bm{\delta}, \varepsilon}{\mathop{\min}}&\sum_{i=1}^n\varepsilon_i^2 &{\quad}&\\
    \mbox{\textit{s.t.}}\quad 
    &  \bm{\gamma}_i^{'}\bm{y}_i = \alpha_i + \bm{\beta}_i^{'}\bm{x}_i + \bm{\delta}_i^{'}\bm{b}_i - \varepsilon_i &{\quad}& \forall i  \notag \\
    &  \alpha_i + \bm{\beta}_i^{'}\bm{x}_i + \delta_i^{'}\bm{b}_i -\bm{\gamma}_i^{'}\bm{y}_i \le \alpha_j + \bm{\beta}_j^{'}\bm{x}_i + \delta_j^{'}\bm{b}_i -\bm{\gamma}_j^{'}\bm{y}_i &{\quad}&  \forall i, j ,\; \text{and} \; i \neq j\notag\\
    &  \bm{\gamma}_i^{'} g^{y}  + \bm{\beta}_i^{'} g^{x} + \bm{\delta}_i^{'}g^{b} = 1  &{\quad}& \forall i  \notag \\
    &  \bm{\beta}_i \ge \bm{0}, \bm{\delta}_i \ge \bm{0}, \bm{\gamma}_i \ge \bm{0} &{\quad}&  \forall i  \notag
\end{alignat}
where the coefficients $\bm{\delta_i}$ denote marginal effects of undesirable outputs to the DDF.

We can also model the undesirable outputs ($\bm{b}$) in the framework of DDF via the module \code{CNLSDDF} (Line 5; cf., the  example above). The estimated coefficients can be displayed by using the function \code{.display_delta()}.
\begin{CodeChunk}
\begin{CodeInput}
>>> from pystoned import CNLSDDF
>>> from pystoned.constant import FUN_PROD, OPT_LOCAL
>>> from pystoned import dataset as dataset
>>> data = dataset.load_GHG_abatement_cost()
>>> model = CNLSDDF.CNLSDDF(y=data.y, x=data.x, b=data.b, 
...     fun=FUN_PROD, gx=[0.0, 0.0], gb=-1.0, gy=1.0)
>>> model.optimize(OPT_LOCAL)
>>> 
>>> model.display_delta()
\end{CodeInput}
\begin{CodeOutput}
delta : delta
    Size=168, Index=delta_index
    Key      : Lower : Value               : Upper : Fixed : Stale : Domain
      (0, 0) :   0.0 :  27.122300643881694 :  None : False : False :  Reals
      (1, 0) :   0.0 :  31.329802702797316 :  None : False : False :  Reals
        .
        .
    (167, 0) :   0.0 :    6.10497985046203 :  None : False : False :  Reals
\end{CodeOutput}
\end{CodeChunk}

\subsubsection{CQR/CER with multiple outputs}

Similar to CNLS with DDF, we present another two approaches integrating DDF to convex quantile/expectile regression, and we also consider modeling the undesirable outputs. 
\begin{itemize}
    \item without undesirable outputs
    \begin{enumerate}[label= \arabic*), leftmargin=0.5cm]
        \setlength{\itemsep}{1pt}
        \setlength{\parskip}{0pt}
        \setlength{\parsep}{0pt}
        \item CQR-DDF model
            \begin{alignat}{2}
            \underset{\alpha,\bm{\beta},\bm{\gamma},\varepsilon^{+},\varepsilon^{-}}{\mathop{\min }}&\,
            \tau \sum\limits_{i=1}^{n}{\varepsilon _{i}^{+}}+(1-\tau )\sum\limits_{i=1}^{n}{\varepsilon _{i}^{-}}  &{\quad}& \\ 
            \mbox{\textit{s.t.}}\quad 
            &  \bm{\gamma}_i^{'}\bm{y}_i = \alpha_i + \bm{\beta}_i^{'}\bm{x}_i + \varepsilon^+_i - \varepsilon^-_i &{\quad}& \forall i \notag \\
            &  \alpha_i + \bm{\beta}_i^{'}\bm{x}_i -\bm{\gamma}_i^{'}\bm{y}_i \le \alpha_j + \bm{\beta}_j^{'}\bm{x}_i -\bm{\gamma}_j^{'}\bm{y}_i &{\quad}&  \forall i, j ,\; \text{and} \; i \neq j\notag\\
            &  \bm{\gamma}_i^{'} g^{y}  + \bm{\beta}_i^{'} g^{x}  = 1  &{\quad}& \forall i \notag \\ 
            &  \bm{\beta}_i \ge \bm{0} , \bm{\gamma}_i \ge \bm{0} &{\quad}&  \forall i \notag \\
            & \varepsilon _i^{+}\ge 0,\ \varepsilon_i^{-} \ge 0 &{\quad}& \forall i \notag
            \end{alignat}
        \item CER-DDF model
            \begin{alignat}{2}
            \underset{\alpha,\bm{\beta},\bm{\gamma},\varepsilon^{+},\varepsilon^{-}}{\mathop{\min}}&\,
            \tilde{\tau} \sum\limits_{i=1}^n(\varepsilon _i^{+})^2+(1-\tilde{\tau} )\sum\limits_{i=1}^n(\varepsilon_i^{-})^2   &{\quad}&  \\ 
            \mbox{\textit{s.t.}}\quad 
            &  \bm{\gamma}_i^{'}\bm{y}_i = \alpha_i + \bm{\beta}_i^{'}\bm{x}_i + \varepsilon^+_i - \varepsilon^-_i &{\quad}& \forall i \notag \\
            &  \alpha_i + \bm{\beta}_i^{'}\bm{x}_i -\bm{\gamma}_i^{'}\bm{y}_i \le \alpha_j + \bm{\beta}_j^{'}\bm{x}_i -\bm{\gamma}_j^{'}\bm{y}_i &{\quad}&  \forall i, j ,\; \text{and} \; i \neq j\notag\\
            &  \bm{\gamma}_i^{'} g^{y}  + \bm{\beta}_i^{'} g^{x}  = 1  &{\quad}& \forall i \notag \\ 
            &  \bm{\beta}_i \ge \bm{0} , \bm{\gamma}_i \ge \bm{0} &{\quad}& \forall i  \notag \\
            & \varepsilon _i^{+}\ge 0,\ \varepsilon_i^{-} \ge 0 &{\quad}& \forall i \notag
            \end{alignat}
    \end{enumerate}
\end{itemize}
Similar to the module \code{CNLSDDF}, the module \code{CQERDDF} uses the same parameters and is applied to estimate the CQR/CER-DDF model. The results can be reported using the functions that have been introduced in module \code{CQER.CQR(y, x, ...)/.CER(y,x,...)}. For instance, the functions \code{.display_positive_residual()} and \code{.display_negative_residual()} are used to display the estimated residuals, respectively (Lines 8 and 9).
\begin{CodeChunk}
\begin{CodeInput}
>>> from pystoned import CQERDDF
>>> from pystoned.constant import FUN_PROD, OPT_LOCAL
>>> from pystoned import dataset as dataset
>>> data = dataset.load_Finnish_electricity_firm(x_select=['OPEX', 'CAPEX'],
...     y_select=['Energy', 'Length', 'Customers'])
>>> model = CQERDDF.CQRDDF(y=data.y, x=data.x, b=None, tau=0.9, 
...     fun = FUN_PROD, gx= [1.0, 0.0], gb=None, gy= [0.0, 0.0, 0.0])
>>> model.optimize(OPT_LOCAL)
>>> 
>>> model.display_positive_residual()
>>> model.display_negative_residual()
\end{CodeInput}
\begin{CodeOutput}
epsilon_plus : positive error term
    Size=89, Index=I
    Key : Lower : Value              : Upper : Fixed : Stale : Domain
      0 :   0.0 :                0.0 :  None : False : False :  Reals
      1 :   0.0 :                0.0 :  None : False : False :  Reals
      .
      .
     88 :   0.0 :                0.0 :  None : False : False :  Reals
\end{CodeOutput}     
\begin{CodeOutput}
epsilon_minus : negative error term
    Size=89, Index=I
    Key : Lower : Value              : Upper : Fixed : Stale : Domain
      0 :   0.0 : 119.91091883695128 :  None : False : False :  Reals
      1 :   0.0 :               -0.0 :  None : False : False :  Reals
      .
      .
     88 :   0.0 : 224.05081364484738 :  None : False : False :  Reals
\end{CodeOutput}
\end{CodeChunk}

\begin{itemize}
    \item with undesirable outputs
    \begin{enumerate}[label= \arabic*), leftmargin=0.5cm]
        \setlength{\itemsep}{1pt}
        \setlength{\parskip}{0pt}
        \setlength{\parsep}{0pt}
        \item CQR-DDF-b model
        \begin{alignat}{2}
        \underset{\alpha,\bm{\beta},\bm{\gamma},\bm{\delta}, \varepsilon^{+},\varepsilon^{-}}{\mathop{\min }}&\,
        \tau \sum\limits_{i=1}^{n}{\varepsilon _{i}^{+}}+(1-\tau )\sum\limits_{i=1}^{n}{\varepsilon _{i}^{-}}  &{\quad}& \\ 
        \mbox{\textit{s.t.}}\quad 
        &  \bm{\gamma}_i^{'}\bm{y_i} = \alpha_i + \bm{\beta}_i^{'}\bm{x_i} + \bm{\delta}_i^{'}\bm{b}_i + \varepsilon^+_i - \varepsilon^-_i &{\quad}& \forall i \notag \\
        &  \alpha_i + \bm{\beta}_i^{'}\bm{x_i} + \bm{\delta}_i^{'}\bm{b}_i -\bm{\gamma}_i^{'}\bm{y_i} \le \alpha_j + \bm{\beta}_j^{'}\bm{x_i} + \bm{\delta}_j^{'}\bm{b}_i -\bm{\gamma}_j^{'}\bm{y_i} &{\quad}& \forall i, j ,\; \text{and} \; i \neq j\notag\\
        &  \bm{\gamma}_i^{'} g^{y}  + \bm{\beta}_i^{'} g^{x} + \bm{\delta}_i^{'}g^{b} = 1 &{\quad}& \forall i \notag \\ 
        &  \bm{\beta}_i \ge \bm{0}, \delta_i \ge \bm{0}, \bm{\gamma}_i \ge \bm{0} &{\quad}&  \forall i \notag \\
        & \varepsilon _i^{+}\ge 0,\ \varepsilon_i^{-} \ge 0 &{\quad}& \forall i \notag
        \end{alignat}
        \item CER-DDF-b model   
        \begin{alignat}{2}
        \underset{\alpha,\bm{\beta}, \bm{\gamma},\bm{\delta}, \varepsilon^{+},\varepsilon^{-}}{\mathop{\min}}&\,
        \tilde{\tau} \sum\limits_{i=1}^n(\varepsilon _i^{+})^2+(1-\tilde{\tau} )\sum\limits_{i=1}^n(\varepsilon_i^{-})^2    &{\quad}&  \\ 
        \mbox{\textit{s.t.}}\quad 
        &  \bm{\gamma}_i^{'}\bm{y_i} = \alpha_i + \bm{\beta}_i^{'}\bm{x_i} + \bm{\delta}_i^{'}\bm{b}_i + \varepsilon^+_i - \varepsilon^-_i  &{\quad}& \forall i \notag \\
        &  \alpha_i + \bm{\beta}_i^{'}\bm{x_i} + \bm{\delta}_i^{'}\bm{b}_i -\bm{\gamma}_i^{'}\bm{y_i} \le \alpha_j + \bm{\beta}_j^{'}\bm{x_i} + \bm{\delta}_j^{'}\bm{b}_i -\bm{\gamma}_j^{'}\bm{y_i}  &{\quad}&  \forall i, j ,\; \text{and} \; i \neq j\notag\\
        &  \bm{\gamma}_i^{'} g^{y}  + \bm{\beta}_i^{'} g^{x} + \bm{\delta}_i^{'}g^{b} = 1   &{\quad}& \forall i \notag \\ 
        &  \bm{\beta}_i \ge \bm{0}, \delta_i \ge \bm{0}, \bm{\gamma}_i \ge \bm{0}  &{\quad}&  \forall i  \notag \\
        & \varepsilon _i^{+}\ge 0,\ \varepsilon_i^{-} \ge 0  &{\quad}& \forall i \notag
        \end{alignat}
    \end{enumerate}
\end{itemize}
The following code shows how to solve the CQR-DDF model when the data include the undesirable output (i.e., GHG emissions). 
\begin{CodeChunk}
\begin{CodeInput}
>>> from pystoned import CQERDDF
>>> from pystoned.constant import FUN_PROD, OPT_LOCAL
>>> from pystoned import dataset as dataset
>>> data = dataset.load_GHG_abatement_cost()
>>> model = CQERDDF.CQRDDF(y=data.y, x=data.x, b=data.b, tau=0.9, 
...     fun = FUN_PROD, gx= [0.0, 0.0], gb=[-1], gy=[1])
>>> model.optimize(OPT_LOCAL)
>>> 
>>> model.display_delta()
\end{CodeInput}
\begin{CodeOutput}
delta : delta
    Size=168, Index=delta_index
    Key      : Lower : Value                   : Upper : Fixed : Stale : Domain
      (0, 0) :   0.0 : -3.1705034323517883e-15 :  None : False : False :  Reals
      (1, 0) :   0.0 :  3.2421934988654777e-15 :  None : False : False :  Reals
        .
        .
    (167, 0) :   0.0 :      0.1270833999407421 :  None : False : False :  Reals
\end{CodeOutput}
\end{CodeChunk}

\subsection{Relaxing convexity}
\subsubsection{Isotonic CNLS}

This section introduces a variant of the CNLS estimator, Isotonic CNLS that relies only on the monotonicity assumption. To relax the concavity assumption in CNLS estimation (i.e., estimating a production function), we have to rephrase the Afriat inequality constraint in Problem \eqref{eq:eq2}. 

Define a binary matrix $\bm{P}=\big[p_{ij} \big]_{n \times n}$ to represent isotonicity (\citeauthor{Keshvari2013}, \citeyear{Keshvari2013}) as follows.
\begin{alignat*}{2}
    p_{ij} = 
    \begin{cases} 
        1   & \text{if } x_i \preccurlyeq x_j \\
        0   & \text{otherwise}
    \end{cases}
\end{alignat*}

Applying the enumeration method to define the elements of matrix $\bm{P}$, we then solve the following QP problem
\begin{alignat}{2}
    \underset{\alpha, \bm{\beta}, \varepsilon} \min &\sum_{i=1}^n\varepsilon_i^2 &{\quad}& \\
    \mbox{\textit{s.t.}}\quad 
    &  y_i = \alpha_i + \bm{\beta}_i^{'}\bm{x}_i + \varepsilon_i &{\quad}& \forall i \notag\\
    &  p_{ij}(\alpha_i + \bm{\beta}_i^{'}\bm{x}_i) \le p_{ij}(\alpha_j + \bm{\beta}_j^{'}\bm{x}_i)  &{\quad}&  \forall i, j ,\; \text{and} \; i \neq j\notag\\
    &  \bm{\beta}_i \ge \bm{0}  &{\quad}&  \forall i \notag
\end{alignat}
Note that the concavity constraints between units $i$ and $j$ are relaxed by the matrix $P_{ij}=0$. If the $P_{ij}=1$ for all $i$ and $j$, then the above QP problem (i.e., ICNLS problem) reduces to the CNLS problem.

To calculate the monotonic models, the \pkg{pyStoNED} package provides the modules \code{ICNLS(y, x, ...)} and \code{ICQER(y, x, ...)}, of which inherit the parameter settings from the modules \code{CNLS(y, x, ...)} and \code{CQER(y, x, ...)}. Therefore, the usages of \code{CNLS()} and \code{CQER()} are similar to those of \code{CNLS()} and \code{CQER()}. Note that the matrix $\bm{P}$ is added in the module as the internal function. The readers could check the source code from the GitHub repository. 
\begin{CodeChunk}
\begin{CodeInput}
>>> from pystoned import ICNLS
>>> from pystoned.constant import CET_ADDI, FUN_PROD, OPT_LOCAL, RTS_VRS
>>> from pystoned.dataset import load_Finnish_electricity_firm
>>> data=load_Finnish_electricity_firm(x_select=['OPEX','CAPEX'], 
...     y_select=['Energy'])
>>> model=ICNLS.ICNLS(y=data.y, x=data.x, z=None, 
...     cet=CET_ADDI, fun=FUN_PROD, rts=RTS_VRS)
>>> model.optimize(OPT_LOCAL)
>>> 
>>> model.display_residual()
\end{CodeInput}
\begin{CodeOutput}
epsilon : residual
    Size=89, Index=I
    Key : Lower : Value                   : Upper : Fixed : Stale : Domain
      0 :  None :   0.0006849098000571985 :  None : False : False :  Reals
      1 :  None :      -3.833461767868812 :  None : False : False :  Reals
      .
      .
     88 :  None :      7.8890389440795445 :  None : False : False :  Reals
\end{CodeOutput}
\end{CodeChunk}

\subsubsection{Isotonic CQR/CER}

Similar to ICNLS, the Isotonic CQR and CER approaches are defined as follows:
\begin{itemize}
    \item ICQR estimator
    \begin{alignat}{2}
    \underset{\alpha, \bm{\beta}, \varepsilon^{+},\varepsilon^{-}}{\mathop{\min }}&\,
    \tau \sum\limits_{i=1}^{n}{\varepsilon _{i}^{+}}+(1-\tau )\sum\limits_{i=1}^{n}{\varepsilon _{i}^{-}}  &{\quad}&  \\ 
    \mbox{\textit{s.t.}}\quad 
    & y_i=\mathbf{\alpha}_i+ \bm{\beta}_i^{'}\bm{x}_i+\varepsilon _i^{+}-\varepsilon _i^{-} &{\quad}&  \forall i  \notag \\
    & p_{ij}(\mathbf{\alpha}_i+\bm{\beta}_{i}^{'}\bm{x}_i) \le p_{ij}(\mathbf{\alpha}_j+\bm{\beta}_j^{'}\bm{x}_i)  &{\quad}&  \forall i, j ,\; \text{and} \; i \neq j\notag\\
    & \bm{\beta}_i\ge \bm{0} &{\quad}&  \forall i   \notag \\
    & \varepsilon _i^{+}\ge 0,\ \varepsilon_i^{-} \ge 0 &{\quad}&  \forall i  \notag
    \end{alignat}
\end{itemize}
\begin{CodeChunk}
\begin{CodeInput}
>>> from pystoned import ICQER
>>> from pystoned.constant import CET_ADDI, FUN_PROD, OPT_LOCAL, RTS_VRS
>>> from pystoned.dataset import load_Finnish_electricity_firm
>>> data=load_Finnish_electricity_firm(x_select=['OPEX','CAPEX'], 
...     y_select=['Energy'])
>>> model = ICQER.ICQR(y=data.y, x=data.x, tau = 0.9, z=None, 
...     cet=CET_ADDI, fun=FUN_PROD, rts=RTS_VRS)
>>> model.optimize(OPT_LOCAL)
>>> 
>>> model.display_residual()
\end{CodeInput}
\begin{CodeOutput}
epsilon : error term
    Size=89, Index=I
    Key : Lower : Value               : Upper : Fixed : Stale : Domain
      0 :  None : -2.0000000585763473 :  None : False : False :  Reals
      1 :  None : -15.000000004675286 :  None : False : False :  Reals
      .
      .
     88 :  None :  -25.00000000488575 :  None : False : False :  Reals
\end{CodeOutput}
\end{CodeChunk}

\begin{itemize}
    \item ICER estimator
    \begin{alignat}{2}
    \underset{\alpha, \bm{\beta}, \varepsilon^{+},\varepsilon^{-}}{\mathop{\min}}&\,
    \tilde{\tau} \sum\limits_{i=1}^n(\varepsilon _i^{+})^2+(1-\tilde{\tau} )\sum\limits_{i=1}^n(\varepsilon_i^{-})^2   &{\quad}&   \\ 
    \mbox{\textit{s.t.}}\quad 
    & y_i=\mathbf{\alpha}_i+ \bm{\beta}_i^{'}\bm{x}_i+\varepsilon _i^{+}-\varepsilon_i^{-} &{\quad}& \forall i  \notag \\
    & p_{ij}(\mathbf{\alpha}_i+\bm{\beta}_{i}^{'}\bm{x}_i) \le p_{ij}(\mathbf{\alpha}_j+\bm{\beta}_j^{'}\bm{x}_i)  &{\quad}&  \forall i, j ,\; \text{and} \; i \neq j\notag\\
    & \bm{\beta}_i\ge \bm{0}  &{\quad}&  \forall i  \notag \\
    & \varepsilon _i^{+}\ge 0,\ \varepsilon_i^{-} \ge 0  &{\quad}&  \forall i  \notag
    \end{alignat}
\end{itemize}
\begin{CodeChunk}
\begin{CodeInput}
>>> from pystoned import ICQER
>>> from pystoned.constant import CET_ADDI, FUN_PROD, OPT_LOCAL, RTS_VRS
>>> from pystoned.dataset import load_Finnish_electricity_firm
>>> data=load_Finnish_electricity_firm(x_select=['OPEX','CAPEX'], 
...     y_select=['Energy'])
>>> model=ICQER.ICER(y=data.y, x=data.x, tau=0.9, z=None, 
...     cet=CET_ADDI, fun=FUN_PROD, rts=RTS_VRS)
>>> model.optimize(OPT_LOCAL)
>>> 
>>> model.display_residual()
\end{CodeInput}
\begin{CodeOutput}
epsilon : error term
    Size=89, Index=I
    Key : Lower : Value                : Upper : Fixed : Stale : Domain
      0 :  None :  -17.735777393077573 :  None : False : False :  Reals
      1 :  None :  -30.733601094693068 :  None : False : False :  Reals
      .
      .
     88 :  None :   -9.466741932963199 :  None : False : False :  Reals
\end{CodeOutput}
\end{CodeChunk}

\section{Stochastic Nonparametric Envelopment of Data} \label{sec:stoned}

Combining virtues of SFA and DEA in a unified framework, StoNED (\citeauthor{Kuosmanen2006}, \citeyear{Kuosmanen2006}; \citeauthor{Kuosmanen2012c}, \citeyear{Kuosmanen2012c}) uses a composed error term to model both inefficiency $u$ and noise $v$ without assuming a functional form of $f$. Analogous to the COLS/C$^2$NLS estimators, the StoNED estimator consists of the following four steps:
\begin{itemize}
    \item Step 1: Estimating the conditional mean $\E[y_i \,| \, \bm{x}_i]$ using CNLS estimator
    \item Step 2: Estimating the expected inefficiency $\mu$ based on the residual $\varepsilon_i^{CNLS}$
    \item Step 3: Estimating the StoNED frontier $\hat{f}^{StoNED}$ based on the $\hat{\mu}$
    \item Step 4: Estimating firm-specific inefficiencies $\E[u_i \mid \varepsilon_i^{CNLS}]$
\end{itemize}

Besides the CNLS estimator, we can apply other convex regression approaches such as ICNLS and CNLS-DDF to estimate the conditional mean in the first step (see \citeauthor{Keshvari2013}, \citeyear{Keshvari2013}; \citeauthor{Kuosmanen2017a}, \citeyear{Kuosmanen2017a}). However, the quantile and expectile related estimators introduced in Section \ref{convexR} can not be integrated into the StoNED framework at present. 

\subsection{Estimating the expected inefficiency}

After obtaining the residuals (e.g., $\hat{\varepsilon}_i^{CNLS}$) from the convex regression approaches, one can estimate the expected value of the inefficiency term $\mu = E(u_i)$. In practice, three commonly used methods are available to estimate the expected inefficiency $\mu$: method of moments (\citeauthor{Aigner1977}, \citeyear{Aigner1977}), quasi-likelihood estimation (\citeauthor{Fan1996}, \citeyear{Fan1996}), and kernel deconvolution estimation (\citeauthor{Hall2002}, \citeyear{Hall2002}). We next briefly review these three approaches and demonstrate the application of \pkg{pyStoNED}; see more detailed theoretical introduction in \cite{Kuosmanen2015d}.

\subsubsection{Method of moments}

The method of moments requires some additional parametric distributional assumptions. Following \citet{Kuosmanen2015d}, under the maintained assumptions of half-normal inefficiency (i.e., $u_i \sim N^+(0, \sigma_u^2)$) and normal noise (i.e., $v_i \sim N(0, \sigma_v^2)$), the second and third central moments of the composite error (i.e., $\varepsilon_i$) distribution are given by
\begin{align}
        M_2 &= \bigg[\frac{\pi-2}{\pi}\bigg] \sigma_u^2 + \sigma_v^2  \notag \\
        M_3 &= \bigg(\sqrt{\frac{2}{\pi}}\bigg)\bigg[1-\frac{4}{\pi}\bigg]\sigma_u^2 \notag
\end{align}

The second and third central moments can be estimated by using the CNLS residuals, i.e.,  $\hat{\varepsilon}_i^{CNLS}$:
    
\begin{align}
    \hat{M_2} &= \sum_{i=1}^{n}(\hat{\varepsilon}_i-\bar{\varepsilon})^{2}/n  \notag \\
    \hat{M_3} &= \sum_{i=1}^{n}(\hat{\varepsilon}_i-\bar{\varepsilon})^{3}/n  \notag
\end{align}

Note that the third moment $M_3$ (which measures the skewness of the distribution) only depends on the standard deviation parameter $\sigma_u$ of the inefficiency distribution. Thus, given the estimated $\hat{M}_3$ (which should be positive in the case of a cost frontier), we can estimate the parameters $\sigma_u$ and $\sigma_v$ by
\begin{align}
    \hat{\sigma}_u &= \sqrt[3]{\frac{\hat{M_3}}{\bigg(\sqrt{\frac{2}{\pi}}\bigg)\bigg[1-\frac{4}{\pi}\bigg]}} \notag \\
    \hat{\sigma}_v &= \sqrt[2]{\hat{M_2}-\bigg[\frac{\pi-2}{\pi}\bigg] \hat{\sigma}_u^2 } \notag
\end{align}

To estimate the unconditional expected inefficiency $\mu$ and firm-specific inefficiencies, we provide a module \code{StoNED()} that inherits the parameter setting from the module \code{CNLS()}. We estimate the conditional expected inefficiency through the presented example below. Following the calculation procedure of StoNED estimator, we first utilize the module \code{CNLS()} to estimate the conditional mean $\E[y_i \,| \, \bm{x}_i]$ (i.e., Lines 5-6), then apply the module \code{StoNED(model)}, where the parameter \code{model} is defined as the name of CNLS model (cf. footnote \ref{fn:fn1}), to decompose the estimated residuals (i.e., Line 8). We develop and import the parameter \code{RED_MOM} (i.e., Line 3) to decompose the residuals by using the method of moment. We finally resort to the function \code{.get_unconditional_expected_inefficiency(RED_MOM))} included in the module \code{StoNED()} to retrieve the expected inefficiency $\hat{\mu}$.
\begin{CodeChunk}
\begin{CodeInput}
>>> from pystoned import CNLS, StoNED
>>> from pystoned.dataset import load_Finnish_electricity_firm
>>> from pystoned.constant import CET_MULT, FUN_COST, RTS_VRS, RED_MOM
>>> data=load_Finnish_electricity_firm(x_select=['Energy','Length','Customers'],
...     y_select=['TOTEX'])
>>> model=CNLS.CNLS(data.y, data.x, z=None, 
...     cet=CET_MULT, fun=FUN_COST, rts=RTS_VRS)
>>> model.optimize('email@address')
>>> 
>>> rd = StoNED.StoNED(model)
>>> print(rd.get_unconditional_expected_inefficiency(RED_MOM))
\end{CodeInput}
\begin{CodeOutput}
0.028561358246550088
\end{CodeOutput}
\end{CodeChunk}

\subsubsection{Quassi-likelihood estimation}

Quasi-likelihood approach is an alternative approach to decomposing $\sigma_u$ and $\sigma_v$ suggested by \citet{Fan1996}. Given the shape of the CNLS curve, we apply the standard maximum likelihood method to estimate the parameters $\sigma_u$ and $\sigma_v$. The quasi-likelihood function is formulated as 
\begin{align}
    \ln L(\lambda) & = -n\ln(\hat{\sigma}) + \sum \ln\Phi\bigg[\frac{-\hat{\varepsilon}_i \lambda}{\hat{\sigma}}\bigg] - \frac{1}{2\hat{\sigma}^2}\sum\hat{\varepsilon}_i^2  \notag
\end{align}
where
\begin{align}
    \hat{\varepsilon}_i &= \hat{\varepsilon}_i^{CNLS}-(\sqrt{2}\lambda\hat{\sigma})/[\pi(1+\lambda^2)]^{1/2}   \notag  \\
    \hat{\sigma} &= \Bigg\{\frac{1}{n}\sum(\hat{\varepsilon}_i^{CNLS})^2 / \bigg[1 - \frac{2\lambda^2}{\pi(1+\lambda^2)}\bigg]  \Bigg\}.   \notag
\end{align}
Note that the quasi-likelihood function only consists of a single parameter $\lambda$ (i.e., the signal-to-noise ratio $\lambda = \sigma_u/\sigma_v$).\footnote{Note the notation difference between signal-to-noise ratio $\lambda$ and the marginal effect of contextual variable $\lambda$ in Problems \eqref{eq:eq10} and \eqref{eq:eq11}.
}
The symbol $\Phi$ represents the cumulative distribution function of the standard normal distribution. In the \pkg{pyStoNED} package, we use the Broyden–Fletcher–Goldfarb–Shanno (BFGS) algorithm provided by \pkg{SciPy} to estimate the maximum likelihood function.

Since we apply the quasi-likelihood estimation to decompose the residuals, we import the parameter \code{RED_QLE} in the module \code{StoNED()} (see Line 3). 
\begin{CodeChunk}
\begin{CodeInput}
>>> from pystoned import CNLS, StoNED
>>> from pystoned.dataset import load_Finnish_electricity_firm
>>> from pystoned.constant import CET_MULT, FUN_COST, RTS_VRS, RED_QLE
>>> data=load_Finnish_electricity_firm(x_select=['Energy','Length','Customers'],
...     y_select=['TOTEX'])
>>> model = CNLS.CNLS(data.y, data.x, z=None, 
...     cet=CET_MULT, fun=FUN_COST, rts=RTS_VRS)
>>> model.optimize('email@address')
>>> 
>>> rd = StoNED.StoNED(model)
>>> print(rd.get_unconditional_expected_inefficiency(RED_QLE))
\end{CodeInput}
\begin{CodeOutput}
0.05776223709846952
\end{CodeOutput}
\end{CodeChunk}

\subsubsection{Kernel deconvolution estimation}

While the method of moments and quasi-likelihood approaches require additional distributional assumptions for the inefficiency and noise terms, an alternative nonparametric estimation of the expected inefficiency $\mu$ is available by applying nonparametric kernel deconvolution proposed by \citet{Hall2002}. Note that the residual $\hat{\varepsilon}_i^{CNLS}$ is a consistent estimator of $e^o = \varepsilon_i + \mu$ (for production model). Following \cite{Kuosmanen2017a}, the density function of ${e^o}$ is
\begin{align}
    \hat{f}_{e^o}(z) = (nh)^{-1} \sum_{i=1}^{n}K\bigg(\frac{z-e_i^o}{h} \bigg), \notag
\end{align}
where $K(\cdot)$ is a compactly supported kernel, and $h$ is a bandwidth. \citet{Hall2002} show that the first derivative of the density function of the composite error term ($f_\varepsilon^{'}$) is proportional to that of the inefficiency term ($f_u^{'}$) in the neighborhood of $\mu$. Therefore, a nonparametric estimator of expected inefficiency $\mu$ is obtained as
\begin{align}
    \hat{\mu} = \arg \max_{z \in C}(\hat{f^{'}}_{e^o}(z)), \notag
\end{align}
where $C$ is a closed interval in the right tail of $f_{e^o}$.

We then import the parameter \code{RED_KDE} to decompose the estimated residuals, i.e., $\hat{\varepsilon}_i^{CNLS}$ (i.e., Line 3).
\begin{CodeChunk}
\begin{CodeInput}
>>> from pystoned import CNLS, StoNED
>>> from pystoned.dataset import load_Finnish_electricity_firm
>>> from pystoned.constant import CET_MULT, FUN_COST, RTS_VRS, RED_KDE
>>> data=load_Finnish_electricity_firm(x_select=['Energy','Length','Customers'],
...     y_select=['TOTEX'])
>>> model = CNLS.CNLS(data.y, data.x, z=None, 
...     cet=CET_MULT, fun=FUN_COST, rts=RTS_VRS)
>>> model.optimize('email@address')
>>> 
>>> rd = StoNED.StoNED(model)
>>> print(rd.get_unconditional_expected_inefficiency(RED_KDE))
\end{CodeInput}
\begin{CodeOutput}
3.1932648065241334
\end{CodeOutput}
\end{CodeChunk}

\subsection{Estimating firm-specific inefficiencies}

After estimating the expected inefficiency $\mu$ using the methods of moment (MOM) or quasi-likelihood estimation (QLE),\footnote{
For the expected inefficiency $\mu$ estimated by kernel deconvolution, \citet{Dai2016} proposes a non-parametric strategy where the Richardson–Lucy blind deconvolution algorithm is used to identify firm-specific inefficiencies. However, the \pkg{pyStoNED} package only supports the parametric estimation of firm-specific inefficiencies due to the fact that the parametric method is more widely used in efficiency analysis literature.
}
we then employ JLMS estimator proposed by \citet{Jondrow1982} to estimate the firm-specific inefficiencies (\citeauthor{Ray2015}, \citeyear{Ray2015}). Under the assumption of a normally distributed error term and a half-normally distributed inefficiency term, JLMS formulates the conditional distribution of inefficiency $u_i$, given $\hat{\varepsilon}_i$, and proposes the inefficiency estimator as the conditional mean $\E[u_i \, | \, \hat{\varepsilon}_i]$. 

Following \citet{Kumbhakar2000}, the conditional expected value of inefficiency $\E[u_i  \, | \, \hat{\varepsilon}_i]$ for production function and cost function are shown as follows, respectively:
    \begin{itemize}
        \item  Production function
\begin{align}
\E[u_i  \, | \, \hat{\varepsilon}_i] = \mu_{*i} + \sigma_* \Bigg[ \frac{\phi(-\mu_{*i}/\sigma_*)}{1-\Phi(-\mu_{*i}/\sigma_*)} \Bigg] = \sigma_* \Bigg[ \frac{\phi(\varepsilon_i \lambda/\sigma)}{1-\Phi(\varepsilon_i \lambda/\sigma)} - \frac{\varepsilon_i \lambda}{\sigma} \Bigg]  \notag
\end{align}
where $\mu_*= -\varepsilon \sigma_u^2/\sigma^2$, $\sigma_*^2 = \sigma_u^2\sigma_v^2/\sigma^2$, $\lambda = \sigma_u/\sigma_v$, and $\sigma^2 = \sigma_u^2 +\sigma_v^2$. The symbol $\phi$ is the standard normal density function, and the symbol $\Phi$ denotes the cumulative distribution function of the standard normal distribution.
       \item Cost function
\begin{align}
    \E[u_i \mid \hat{\varepsilon}_i] = \mu_{*i} + \sigma_* \Bigg[ \frac{\phi(-\mu_{*i}/\sigma_*)}{1-\Phi(-\mu_{*i}/\sigma_*)} \Bigg] = \sigma_* \Bigg[ \frac{\phi(\varepsilon_i \lambda/\sigma)}{1-\Phi(-\varepsilon_i \lambda/\sigma)} + \frac{\varepsilon_i \lambda}{\sigma} \Bigg]  \notag
\end{align}
where $\mu_*= \varepsilon \sigma_u^2/\sigma^2$, $\sigma_*^2 = \sigma_u^2\sigma_v^2/\sigma^2$, $\lambda = \sigma_u/\sigma_v$, and $\sigma^2 = \sigma_u^2 +\sigma_v^2$.
    \end{itemize}

The firm-level technical efficiency (TE) is then measured based on the estimated conditional mean. For other models, the technical efficiency can be estimated as follows. 
\begin{itemize}
    \item Production function
    \begin{itemize}
        \item  Multiplicative model: $\text{TE} = \exp(-\E[u_i \mid  \varepsilon_i])$ 
        \item  Additive model: $\text{TE} = \frac{y - \E[u_i \mid  \varepsilon_i]}{y}$
    \end{itemize}
    \item Cost function
    \begin{itemize}
        \item  Multiplicative model: $\text{TE} = \exp(\E[u_i \mid  \varepsilon_i])$
        \item  Additive model: $\text{TE} = \frac{y+ \E[u_i \mid  \varepsilon_i]}{y}$
    \end{itemize}
\end{itemize}

To calculate the firm-level technical efficiency, we resort to the function \code{.get_technical_\\inefficiency()}. We either set the parameter in function \code{.get_technical_inefficiency()} to \code{RED_MOM} (i.e., using the MOM approach to calculate the efficiency; Line 9) or to \code{RED_QLE} (i.e., using the quassi-likelihood estimation approach).
\begin{CodeChunk}
\begin{CodeInput}
>>> from pystoned import CNLS, StoNED
>>> from pystoned.dataset import load_Finnish_electricity_firm
>>> from pystoned.constant import CET_MULT, FUN_COST, RTS_VRS, RED_MOM
>>> data=load_Finnish_electricity_firm(x_select=['Energy','Length','Customers'],
...     y_select=['TOTEX'])
>>> model = CNLS.CNLS(data.y, data.x, z=None, 
...     cet=CET_MULT, fun=FUN_COST, rts=RTS_VRS)
>>> model.optimize('email@address')
>>> 
>>> rd = StoNED.StoNED(model)
>>> print(rd.get_technical_inefficiency(RED_MOM))  
\end{CodeInput}
\begin{CodeOutput}
[1.02974326 1.029505   1.03407182 1.03011172 1.03285089 1.02624282
 1.02854464 1.03305965 ...        1.02883011 1.03059918 1.03003701]
\end{CodeOutput}
\end{CodeChunk}

Further, we can consider the effect of contextual variables ($\bm{z}$) in the first-step estimation of the StoNED estimator (i.e., Line 5) when calculating the technical efficiency. 
\begin{CodeChunk}
\begin{CodeInput}
>>> from pystoned import CNLS, StoNED
>>> from pystoned.dataset import load_Finnish_electricity_firm
>>> from pystoned.constant import CET_MULT, FUN_COST, RTS_VRS, RED_MOM    
>>> data=load_Finnish_electricity_firm(x_select=['Energy','Length','Customers'],
...     y_select=['TOTEX'], z_select=['PerUndGr'])
>>> model = CNLS.CNLS(y=data.y, x=data.x, z=data.z, 
...     cet=CET_MULT, fun=FUN_COST, rts=RTS_VRS)
>>> model.optimize('email@address')
>>> 
>>> rd = StoNED.StoNED(model)
>>> print(rd.get_technical_inefficiency(RED_MOM))
\end{CodeInput}
\begin{CodeOutput}
[1.03110069 1.02940641 1.02886856 1.02977033 1.03447988 1.02600838
 1.02809928 1.03574493 ...        1.02743418 1.03236693 1.02948783]
 
\end{CodeOutput}
\end{CodeChunk}

\section{CNLS-G Algorithm}\label{sec:cnlsg}

Since convex regression approaches shape the convexity (concavity) of function using the Afrait inequality, the estimation becomes excessively expensive due to the $O(n^2)$ linear constraints. For example, if the data samples have 500 observations, the total number of linear constraints is equal to 250,000. To speed up the computational time, \citet{Lee2013} propose a more efficient generic algorithm, CNLS-G, which uses the relaxed Afriat constraint set and iteratively adds violated constraints to the relaxed model as necessary. See further discussions in \citet{Lee2013}.

To illustrate the CNLS-G algorithm, we follow \citet{Lee2013} to generate the input and output variables. In this section, we assume an additive production function with two-input and one-output, $y=x_1^{0.4} \times x_2^{0.4}+u$. We randomly draw the inputs $x_1$ and $x_2$ from a uniform distribution, $x \sim U[1, 10]$, and the error term $u$ from a normal distribution, $u \sim N(0, 0.7^2)$. Based on these settings, we generate 500 artificial observations, estimate the CNLS problem \eqref{eq:eq2} and the CER problem \eqref{eq:eq8}, and calculate the firm-level technical efficiency using the CNLS-G algorithm.

\subsection{Solving CNLS model}\label{sec:cnls-g}

We first compare the running time of original CNLS and CNLS-G algorithm in the same computation environment.  Line 1 imports the modules \code{CNLSG()} that is designed to perform the CNLS-G algorithm and \code{CNLS()} from the package \pkg{pyStoNED}. Note that the module \code{CNLSG()} has the same parameters as the module \code{CNLS()}. Line 3 imports the \pkg{NumPy} to provide multidimensional array and functions for linear algebra. We use the \pkg{time} to count the running time for CNLS estimation (i.e., Lines 4, 9,  12) and employ the function \code{.get_runningtime()} to directly obtain the running time for CNLS-G algorithm (i.e., Line 16). To replicate the experiment, we set a random seed using the \code{np.random.seed()}. Lines 6-8 generate the variables $\bm{x}$, $u$, and $y$. \code{model1} and \code{model2} are the normal CNLS model and CNLS-G model, respectively. To count the number of constraints included in the CNLS-G algorithm, the module \code{CNLSG()} provides an internal function \code{.get_totalconstr()} (Line 19).
\begin{CodeChunk}
\begin{CodeInput}
>>> from pystoned import CNLSG, CNLS
>>> from pystoned.constant import CET_ADDI, FUN_PROD, OPT_LOCAL, RTS_VRS
>>> import numpy as np
>>> import time
>>> np.random.seed(0)
>>> x=np.random.uniform(low=1, high=10, size=(500, 2))
>>> u=np.random.normal(loc=0, scale=0.7, size=500)
>>> y=x[:, 0]**0.4*x[:, 1]**0.4+u
>>> t1=time.time()
>>> model1=CNLS.CNLS(y, x, z=None, 
...     cet=CET_ADDI, fun=FUN_PROD, rts=RTS_VRS)
>>> model1.optimize(OPT_LOCAL)
>>> CNLS_time = time.time()-t1
>>> model2=CNLSG.CNLSG(y, x, z=None, 
...     cet=CET_ADDI, fun=FUN_PROD, rts=RTS_VRS)
>>> model2.optimize(OPT_LOCAL)
>>> 
>>> print("The running time with algorithm is ", model2.get_runningtime())
>>> print("The running time without algorithm is ", CNLS_time)
>>> 
>>> print("The total number of constraints is ", model2.get_totalconstr())
\end{CodeInput}

\begin{CodeOutput}
The running time with algorithm is  20.35739278793335
The running time without algorithm is  30.252474308013916
The total number of constraints is  9700.0
\end{CodeOutput}
\end{CodeChunk}

\subsection{Solving CER model}

We next demonstrate a CER model solved by the CNLS-G algorithm prepared in module \code{CQERG()}. The other experimental settings are similar to those in Section \ref{sec:cnls-g}. Note that the CQR model can also be solved by the CNLS-G algorithm via the function \code{CQERG.CQRG(y, x, ...)} (cf. Line 14).
\begin{CodeChunk}
\begin{CodeInput}
>>> from pystoned import CQERG, CQER
>>> from pystoned.constant import CET_ADDI, FUN_PROD, OPT_LOCAL, RTS_VRS
>>> import numpy as np
>>> import time
>>> np.random.seed(0)
>>> x=np.random.uniform(low=1, high=10, size=(500, 2))
>>> u=np.random.normal(loc=0, scale=0.7, size=500)
>>> y=x[:, 0]**0.4*x[:, 1]**0.4+u
>>> tau=0.5
>>> t1=time.time()
>>> model1=CQER.CER(y, x, tau, z=None, 
...     cet=CET_ADDI, fun=FUN_PROD, rts=RTS_VRS)
>>> model1.optimize(OPT_LOCAL)
>>> CER_time=time.time()-t1
>>> model2=CQERG.CERG(y, x, tau, z=None, 
...     cet=CET_ADDI, fun=FUN_PROD, rts=RTS_VRS)
>>> model2.optimize(OPT_LOCAL)
>>> 
>>> print("The running time with algorithm is ", model2.get_runningtime())
>>> print("The running time without algorithm is ", CER_time)
>>> print("The total number of constraints in CER model is ", 
...     model2.get_totalconstr())
\end{CodeInput}
\begin{CodeOutput}
The running time with algorithm is  22.845768451690674
The running time without algorithm is  29.961899280548096
The total number of constraints in CER model is  9701.0
\end{CodeOutput}
\end{CodeChunk}

\subsection{Calculating firm-level efficiency}

We can apply the CNLS-G algorithm to calculate the firm-level efficiency more efficiently. In the first step of \code{StoNED()} estimator, we use the module \code{CNLSG(y, x, ...)} as a substitute for the module \code{CNLS(y, x, ...)} (Lines 1 and 5). The other settings are similar to those in Section \ref{sec:stoned}.
\begin{CodeChunk}
\begin{CodeInput}
>>> from pystoned import CNLSG, StoNED
>>> from pystoned.dataset import load_Finnish_electricity_firm
>>> from pystoned.constant import CET_MULT, FUN_COST, RTS_VRS, RED_MOM
>>> data=load_Finnish_electricity_firm(x_select=['Energy','Length','Customers'],
...     y_select=['TOTEX'])
>>> model=CNLSG.CNLSG(data.y, data.x, z=None, 
...     cet=CET_MULT, fun=FUN_COST, rts=RTS_VRS)
>>> model.optimize('email@address')
>>>  
>>> rd = StoNED.StoNED(model)
>>> print(rd.get_technical_inefficiency(RED_MOM))
\end{CodeInput}
\begin{CodeOutput}
[1.09156319 1.08650469 1.14247297 1.09270023 1.12171169 1.06035128
 1.07631848 1.13245426 ...        1.08719327 1.09467879 1.09408226]
\end{CodeOutput}
\end{CodeChunk}

\section{Graphical illustration of estimated function}\label{sec:plot}
To illustrate how the estimated function looks like, the \pkg{pyStoNED} package provides the functions \code{plot.plot2d()} and \code{plot.plot3d()} to plot two- and three-dimensional (2D and 3D) estimated functions. Similar to the usage of the module \code{StoNED(y, x, ...)}, we first estimate the nonparametric regression such as CNLS, CQR, and ICNLS  and then apply the \code{plot} function to draw the figures. In this section, we use the internal data provided with Tim Coelli's Frontier 4.1 to demonstrate the plotting process.

\subsection{One-input and One-output}

\begin{figure}[!htbp]
    \vspace{-2em}
    \centering
    \subfloat[\centering CNLS model]{{\includegraphics[width=8.2cm]{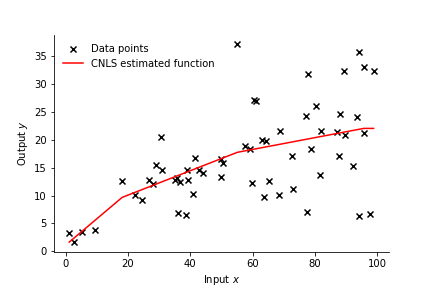} }}%
    \subfloat[\centering CQR model ($\tau=0.5$)]{{\includegraphics[width=8.2cm]{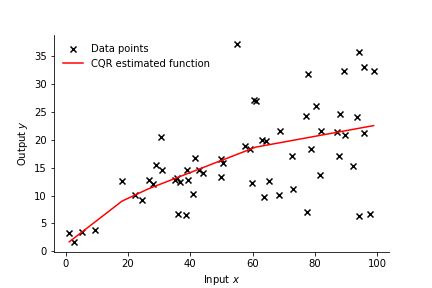} }}%
    \caption{Estimated function by CNLS/CQR model.}%
    \label{fig:fig1}%
\end{figure}

In the one-input and one-output case, we present two different estimated functions: the CNLS function and the CQR function. Therefore, we import the modules \code{CNLS} and \code{CQER} in Line 1 as the estimators, the plotting module \code{plot2d} (Line 2), and the example dataset (Line 4). Lines 6 and 7 and Lines 10 and 11 define and solve the CNLS and CQR models. Lines 8 and 12 are used to plot the estimated function. There are four parameters in the module \code{plot2d(...)}: the first is the model's name (see footnote \ref{fn:fn1}), the second parameter \code{x\_select} defines which is the selected input $x$, the third and last are the given names of the legend and generated picture, respectively. 
\begin{CodeChunk}
\begin{CodeInput}
>>> from pystoned import CNLS, CQER
>>> from pystoned.plot import plot2d
>>> from pystoned.constant import CET_ADDI, FUN_PROD, OPT_LOCAL, RTS_VRS
>>> from pystoned.dataset import load_Tim_Coelli_frontier
>>> data=load_Tim_Coelli_frontier(x_select=['labour'],y_select=['output'])
>>> CNLS_model=CNLS.CNLS(y=data.y, x=data.x, z=None, 
...     cet=CET_ADDI, fun=FUN_PROD, rts=RTS_VRS)
>>> CNLS_model.optimize(OPT_LOCAL)
>>> plot2d(CNLS_model, x_select=0, label_name="CNLS estimated function", 
...     fig_name="CNLS_2d")
>>> 
>>> CQR_model=CQER.CQR(y=data.y, x=data.x, tau=0.5, z=None, 
...     cet=CET_ADDI, fun=FUN_PROD, rts=RTS_VRS)
>>> CQR_model.optimize(OPT_LOCAL)
>>> plot2d(CQR_model, x_select=0, label_name="CQR estimated function",
...     fig_name="CQR_2d")
\end{CodeInput}
\end{CodeChunk}
Figure \ref{fig:fig1} depicts the functions estimated by the CNLS and CQR model (with $\tau=0.5$).

\subsection{Two-input and One-output}

\begin{figure}[!htbp]
    \vspace{-1em}
    \centering
    \includegraphics[width=12cm]{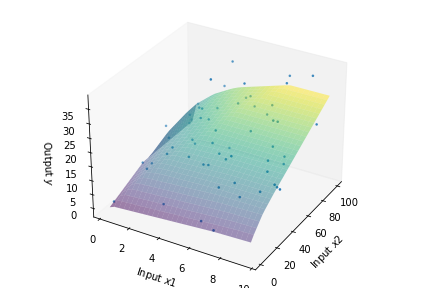}
    \caption{Estimated function by CNLS model.}
    \label{fig:fig2}
\end{figure}

In the two-input and one-output case, we use linear interpolation to obtain the 3D surface due to the fact that CNLS estimates hyperplanes at the observation points.\footnote{
Note that the CNLS estimator includes linear programming as the second stage estimation to find the minimum extrapolated production function (\citeauthor{Kuosmanen2012c}, \citeyear{Kuosmanen2012c}).
}
The function \code{plot3d(model, x_select_1, x_select_2, fig_name=None, line_transparent=False, \\pane_transparent=False)} includes six parameters: the first is the name of the estimated model, the second and third define the selected input, the fourth is the name of generated figure, the last two are the basic settings for the figure (False; default). We import function \code{plot3d(...)} in Line 2 and plot the figure in Line 8. Figure \ref{fig:fig2} presents the estimated 3D estimated function.
\begin{CodeChunk}
\begin{CodeInput}
>>> from pystoned import CNLS
>>> from pystoned.plot import plot3d
>>> from pystoned.constant import CET_ADDI, FUN_PROD, OPT_LOCAL, RTS_VRS
>>> from pystoned.dataset import load_Tim_Coelli_frontier
>>> data=load_Tim_Coelli_frontier(x_select=['capital', 'labour'],
...     y_select=['output'])
>>> CNLS_model=CNLS.CNLS(y=data.y, x=data.x, z=None,
...     cet=CET_ADDI, fun=FUN_PROD, rts=RTS_VRS)
>>> CNLS_model.optimize(OPT_LOCAL)
>>> plot3d(CNLS_model,x_select_1=0, x_select_2=1, fig_name="CNLS_3d")
\end{CodeInput}
\end{CodeChunk}

\section{Conclusions}\label{sec:conc}

Convex regression and related methods provide an appealing way to impose shape constraints implied by the theory without making restrictive functional form assumptions. As these techniques are becoming increasingly popular, there is a great demand for a powerful, reliable, and fully open access computational package. The \proglang{Python} package \pkg{pyStoNED} aims to address this need, providing a comprehensive set of functions for estimating CNLS, StoNED, and their numerous variants. For more information, see its developed repository on GitHub at \url{https://github.com/ds2010/pyStoNED} and documentation at \url{https://pystoned.readthedocs.io}. 

This paper has reviewed the models and specifications currently supported by \pkg{pyStoNED}, and demonstrated its use by empirical examples taken from the literature of productivity and efficiency analysis. All modules are implemented in a fully open access environment. We encourage the users to utilize \pkg{pyStoNED} to further develop their own packages for specific estimation purposes. 

In the future, our plan is to include further extensions to \pkg{pyStoNED} on a continuous basis. One interesting avenue of ongoing research is to include additional penalty terms to the objective function of the CNLS/CQR/CER problems to alleviate overfitting and the curse of dimensionality (e.g.,  \citeauthor{Lee2020}, \citeyear{Lee2020}). More efficient computational algorithms such as the CNLS-A proposed by \citet{Dai2021} are under active development, and will be included in the future editions of \pkg{pyStoNED}.

We hope that the \pkg{pyStoNED} package could contribute to raising the standards of empirical applications of productivity and efficiency analysis, and facilitate more meaningful and relevant applications that influence the managerial and policy decisions. The basic idea of the StoNED approach is to enable applied researchers and practitioners of efficiency analysis to incorporate existing tools and techniques from different domains such as econometrics, statistics, operational research, and machine learning to a logically consistent unified framework, but also to facilitate further methodological development in a multi-disciplinary environment. The purpose of the \pkg{pyStoNED} package is to support this development.

While we have phrased this review and the \pkg{pyStoNED} modules in terms of the cost and production function, most of the modules are readily applicable to nonparametric regression analysis in any other context as well. We hope that the convex regression and related techniques could also prove useful in other application areas where shape constraints play an essential role. For example, optimization behavior also implies specific convexity constraints in the context of consumer demand analysis (see, e.g., \citeauthor{Afriat1967}, \citeyear{Afriat1967}; \citeauthor{Varian1982}, \citeyear{Varian1982}).

\section*{Acknowledgments}
Sheng Dai gratefully acknowledges financial support from the Foundation for Economic Education (Liikesivistysrahasto) [No. 180019, 190073] and the HSE Support Foundation [No. 11–2290].

\bibliography{refs}

\begin{thebibliography}{75}
\newcommand{\enquote}[1]{``#1''}
\providecommand{\natexlab}[1]{#1}
\providecommand{\url}[1]{\texttt{#1}}
\providecommand{\urlprefix}{URL }
\expandafter\ifx\csname urlstyle\endcsname\relax
  \providecommand{\doi}[1]{doi:\discretionary{}{}{}#1}\else
  \providecommand{\doi}{doi:\discretionary{}{}{}\begingroup
  \urlstyle{rm}\Url}\fi
\providecommand{\eprint}[2][]{\url{#2}}

\bibitem[{Afriat(1972)}]{Afriat1972}
Afriat SN (1972).
\newblock \enquote{{Efficiency Estimation of Production Functions}.}
\newblock \emph{International Economic Review}, \textbf{13}, 568--598.
\newblock \doi{10.2307/2525845}.

\bibitem[{Afriat(1967)}]{Afriat1967}
Afriat SN (1967).
\newblock \enquote{{The Construction of Utility Functions from Expenditure
  Data}.}
\newblock \emph{International Economic Review}, \textbf{8}, 67.
\newblock \doi{10.2307/2525382}.

\bibitem[{Aigner \emph{et~al.}(1977)Aigner, Lovell, and Schmidt}]{Aigner1977}
Aigner D, Lovell CK, Schmidt P (1977).
\newblock \enquote{{Formulation and Estimation of Stochastic Frontier
  Production Function Models}.}
\newblock \emph{Journal of Econometrics}, \textbf{6}, 21--37.
\newblock \doi{10.1016/0304-4076(77)90052-5}.

\bibitem[{{\'{A}}lvarez \emph{et~al.}(2020){\'{A}}lvarez, Barbero, and
  Zof{\'{i}}o}]{Alvarez2020}
{\'{A}}lvarez IC, Barbero J, Zof{\'{i}}o JL (2020).
\newblock \enquote{{A Data Envelopment Analysis Toolbox for
  \proglang{MATLAB}}.}
\newblock \emph{Journal of Statistical Software}, \textbf{95}, 1--49.
\newblock \doi{10.18637/jss.v095.i03}.

\bibitem[{Badunenko \emph{et~al.}(2020)Badunenko, Mozharovskyi, and
  Kolomiytseva}]{oleg2020}
Badunenko O, Mozharovskyi P, Kolomiytseva Y (2020).
\newblock \enquote{{\pkg{npsf}: Nonparametric and Stochastic Efficiency and
  Productivity Analysis}.}
\newblock \proglang{R} package version 0.8-0,
  \urlprefix\url{https://CRAN.R-project.org/package=npsf}.

\bibitem[{Barbero and Zofío(2021)}]{deajl}
Barbero J, Zofío J (2021).
\newblock \enquote{\pkg{DataEnvelopmentAnalysis}: A \proglang{Julia} Package
  for Efficiency and Productivity Measurement using Data Envelopment Analysis
  (DEA).}
\newblock \doi{10.5281/zenodo.5037129}.

\bibitem[{Belotti \emph{et~al.}(2013)Belotti, Daidone, Ilardi, and
  Atella}]{belotti2013}
Belotti F, Daidone S, Ilardi G, Atella V (2013).
\newblock \enquote{Stochastic Frontier Analysis using Stata.}
\newblock \emph{The Stata Journal}, \textbf{13}, 719--758.
\newblock \doi{10.1177/1536867X1301300404}.

\bibitem[{Bertsimas and Mundru(2021)}]{Bertsimas2020}
Bertsimas D, Mundru N (2021).
\newblock \enquote{{Sparse convex regression}.}
\newblock \emph{INFORMS Journal on Computing}, \textbf{33}, 262--279.
\newblock \doi{10.1287/ijoc.2020.0954}.

\bibitem[{Bogetoft and Otto(2010)}]{bogetoft2010}
Bogetoft P, Otto L (2010).
\newblock \emph{{Benchmarking With DEA, SFA, and \proglang{R}}}.
\newblock Springer-Verlag, New York.

\bibitem[{Brunk(1955)}]{Brunk1955}
Brunk HD (1955).
\newblock \enquote{{Maximum Likelihood Estimates of Monotone Parameters}.}
\newblock \emph{The Annals of Mathematical Statistics}, \textbf{26}, 607--616.
\newblock \doi{10.1214/aoms/1177728420}.

\bibitem[{Bynum \emph{et~al.}(2021)Bynum, Hackebeil, Hart, Laird, Nicholson,
  Siirola, Watson, and Woodruff}]{bynum2021pyomo}
Bynum ML, Hackebeil GA, Hart WE, Laird CD, Nicholson BL, Siirola JD, Watson JP,
  Woodruff DL (2021).
\newblock \emph{\pkg{Pyomo}--Optimization Modeling in \proglang{Python}},
  volume~67.
\newblock Third edition. Springer, Berlin.

\bibitem[{Byrd \emph{et~al.}(2006)Byrd, Nocedal, and Waltz}]{byrd2006}
Byrd RH, Nocedal J, Waltz RA (2006).
\newblock \enquote{{\pkg{KNITRO}: An Integrated Package for Nonlinear
  Optimization}.}
\newblock In \emph{Large-scale nonlinear optimization}, pp. 35--59. Springer.

\bibitem[{Chambers \emph{et~al.}(1996)Chambers, Chung, and
  F{\"{a}}re}]{Chambers1996}
Chambers RG, Chung Y, F{\"{a}}re R (1996).
\newblock \enquote{{Benefit and Distance Functions}.}
\newblock \emph{Journal of Economic Theory}, \textbf{70}, 407--419.
\newblock \doi{10.1006/jeth.1996.0096}.

\bibitem[{Chambers \emph{et~al.}(1998)Chambers, Chung, and
  F{\"{a}}re}]{Chambers1998b}
Chambers RG, Chung Y, F{\"{a}}re R (1998).
\newblock \enquote{{Profit, Directional Distance Functions, and Nerlovian
  Efficiency}.}
\newblock \emph{Journal of Optimization Theory and Applications}, \textbf{98},
  351--364.
\newblock \doi{10.1023/A:1022637501082}.

\bibitem[{Charnes \emph{et~al.}(1978)Charnes, Cooper, and Rhodes}]{Charnes1978}
Charnes A, Cooper W, Rhodes E (1978).
\newblock \enquote{{Measuring the Efficiency of Decision Making Units}.}
\newblock \emph{European Journal of Operational Research}, \textbf{2},
  429--444.
\newblock \doi{10.1016/0377-2217(78)90138-8}.

\bibitem[{Coelli(1996)}]{coelli1996}
Coelli T (1996).
\newblock \enquote{A Guide to FRONTIER Version 4.1: A Computer Program for
  Stochastic Frontier Production and Cost Function Estimation.}
\newblock \emph{{CEPA} working paper 96/08}, University of New England.

\bibitem[{Coelli and Henningsen(2020)}]{Tim2020}
Coelli T, Henningsen A (2020).
\newblock \enquote{{\pkg{frontier}: Stochastic Frontier Analysis}.}
\newblock \proglang{R} package version 1.1-8,
  \urlprefix\url{https://CRAN.R-project.org/package=frontier}.

\bibitem[{Coelli \emph{et~al.}(2005)Coelli, Rao, O'Donnell, and
  Battese}]{coelli2005}
Coelli T, Rao DSP, O'Donnell CJ, Battese GE (2005).
\newblock \emph{An Introduction to Efficiency and Productivity Analysis}.
\newblock Springer, New York.

\bibitem[{{Cplex, IBM ILOG}(2009)}]{cplex2009}
{Cplex, IBM ILOG} (2009).
\newblock \enquote{{V12. 1: User’s Manual for \pkg{CPLEX}}.}
\newblock \emph{International Business Machines Corporation}, \textbf{46}(53),
  157.

\bibitem[{Dai(2021)}]{Dai2021}
Dai S (2021).
\newblock \enquote{{Variable Selection in Convex Quantile Regression: L1-norm
  or L0-norm Regularization?}}
\newblock \urlprefix\url{https://arxiv.org/abs/2107.03119}.

\bibitem[{Dai \emph{et~al.}(2020)Dai, Zhou, and Kuosmanen}]{Dai2020}
Dai S, Zhou X, Kuosmanen T (2020).
\newblock \enquote{{Forward-looking Assessment of the GHG Abatement Cost:
  Application to China}.}
\newblock \emph{Energy Economics}, \textbf{88}, 104758.
\newblock \doi{10.1016/j.eneco.2020.104758}.

\bibitem[{Dai(2016)}]{Dai2016}
Dai X (2016).
\newblock \enquote{{Non-parametric Efficiency Estimation Using Richardson-Lucy
  Blind Deconvolution}.}
\newblock \emph{European Journal of Operational Research}, \textbf{248},
  731--739.
\newblock \doi{10.1016/j.ejor.2015.08.004}.

\bibitem[{Dakpo \emph{et~al.}(2021)Dakpo, Desjeux, and Latruffe}]{Dakpo2021}
Dakpo KH, Desjeux D, Latruffe L (2021).
\newblock \enquote{{\pkg{sfaR}: Stochastic Frontier Analysis Using
  \proglang{R}}.}
\newblock \proglang{R} package version 0.1-0,
  \urlprefix\url{https://CRAN.R-project.org/package=sfaR}.

\bibitem[{Daouia and Laurent(2013)}]{Daouia2020}
Daouia A, Laurent T (2013).
\newblock \enquote{{\pkg{frontiles}: Partial Frontier Efficiency Analysis}.}
\newblock \proglang{R} package version 1.2,
  \urlprefix\url{https://CRAN.R-project.org/package=frontiles}.

\bibitem[{Fan \emph{et~al.}(1996)Fan, Li, and Weersink}]{Fan1996}
Fan Y, Li Q, Weersink A (1996).
\newblock \enquote{{Semiparametric Estimation of Stochastic Production Frontier
  Models}.}
\newblock \emph{Journal of Business \& Economic Statistics}, \textbf{14},
  460--468.
\newblock \doi{10.2307/1392254}.

\bibitem[{{GAMS Development Corporation}(2013)}]{GamsSoftware2013}
{GAMS Development Corporation} (2013).
\newblock \emph{{General Algebraic Modeling System (\proglang{GAMS}) Release
  24.2.1}}.
\newblock Washington, DC, USA.
\newblock \urlprefix\url{http://www.gams.com}.

\bibitem[{Grenander(1956)}]{Grenander1956}
Grenander U (1956).
\newblock \enquote{{On the theory of mortality measurement}.}
\newblock \emph{Scandinavian Actuarial Journal}, \textbf{1956}(2), 125--153.
\newblock \doi{10.1080/03461238.1956.10414944}.

\bibitem[{Hall and Simar(2002)}]{Hall2002}
Hall P, Simar L (2002).
\newblock \enquote{{Estimating a Change Point, Boundary, or Frontier in the
  Presence of Observation Error}.}
\newblock \emph{Journal of the American Statistical Association}, \textbf{97},
  523--534.
\newblock \doi{10.1198/016214502760047050}.

\bibitem[{Hannah and Dunson(2013)}]{Hannah2013}
Hannah LA, Dunson DB (2013).
\newblock \enquote{{Multivariate convex regression with adaptive
  partitioning}.}
\newblock \emph{Journal of Machine Learning Research}, \textbf{14}, 3153--3188.
\newblock \urlprefix\url{http://jmlr.org/papers/v14/hannah13a.html}.

\bibitem[{Harris \emph{et~al.}(2020)Harris, Millman, van~der Walt, Gommers,
  Virtanen, Cournapeau, Wieser, Taylor, Berg, Smith, Kern, Picus, Hoyer, van
  Kerkwijk, Brett, Haldane, del R{\'{i}}o, Wiebe, Peterson,
  G{\'{e}}rard-Marchant, Sheppard, Reddy, Weckesser, Abbasi, Gohlke, and
  Oliphant}]{harris2020}
Harris CR, Millman KJ, van~der Walt SJ, Gommers R, Virtanen P, Cournapeau D,
  Wieser E, Taylor J, Berg S, Smith NJ, Kern R, Picus M, Hoyer S, van Kerkwijk
  MH, Brett M, Haldane A, del R{\'{i}}o JF, Wiebe M, Peterson P,
  G{\'{e}}rard-Marchant P, Sheppard K, Reddy T, Weckesser W, Abbasi H, Gohlke
  C, Oliphant TE (2020).
\newblock \enquote{Array programming with \pkg{NumPy}.}
\newblock \emph{Nature}, \textbf{585}, 357--362.
\newblock \doi{10.1038/s41586-020-2649-2}.

\bibitem[{Hildreth(1954)}]{Hildreth1954}
Hildreth C (1954).
\newblock \enquote{{Point Estimates of Ordinates of Concave Functions}.}
\newblock \emph{Journal of the American Statistical Association}, \textbf{49},
  598--619.
\newblock \doi{10.2307/2281132}.

\bibitem[{Hunter(2007)}]{Hunter2007}
Hunter JD (2007).
\newblock \enquote{Matplotlib: A 2D Graphics Environment.}
\newblock \emph{Computing in Science \& Engineering}, \textbf{9}, 90--95.
\newblock \doi{10.1109/MCSE.2007.55}.

\bibitem[{Ji and Lee(2010)}]{ji2010}
Ji Yb, Lee C (2010).
\newblock \enquote{Data Envelopment Analysis.}
\newblock \emph{The Stata Journal}, \textbf{10}, 267--280.
\newblock \doi{10.1177/1536867X1001000207}.

\bibitem[{Johnson and Kuosmanen(2011)}]{Johnson2011}
Johnson AL, Kuosmanen T (2011).
\newblock \enquote{{One-stage Estimation of the Effects of Operational
  Conditions and Practices on Productive Performance: Asymptotically Normal and
  Efficient, Root-n Consistent StoNEZD Method}.}
\newblock \emph{Journal of Productivity Analysis}, \textbf{36}, 219--230.
\newblock \doi{s11123-011-0231-5}.

\bibitem[{Johnson and Kuosmanen(2012)}]{Johnson2012a}
Johnson AL, Kuosmanen T (2012).
\newblock \enquote{{One-stage and two-stage DEA estimation of the effects of
  contextual variables}.}
\newblock \emph{European Journal of Operational Research}, \textbf{220},
  559--570.
\newblock \doi{10.1016/j.ejor.2012.01.023}.

\bibitem[{Johnson and Kuosmanen(2015)}]{Ray2015}
Johnson AL, Kuosmanen T (2015).
\newblock \enquote{{An Introduction to CNLS and StoNED Methods for Efficiency
  Analysis: Economic Insights and Computational Aspects}.}
\newblock In SC~Ray, SC~Kumbhakar, P~Dua (eds.), \emph{Benchmarking for
  Performance Evaluation: A Production Frontier Approach}, chapter~3, pp.
  117--186. Springer.

\bibitem[{Jondrow \emph{et~al.}(1982)Jondrow, Lovell, Materov, and
  Schmidt}]{Jondrow1982}
Jondrow J, Lovell CK, Materov IS, Schmidt P (1982).
\newblock \enquote{{On the Estimation of Technical Inefficiency in the
  Stochastic Frontier Production Function Model}.}
\newblock \emph{Journal of Econometrics}, \textbf{19}, 233--238.
\newblock \doi{10.1016/0304-4076(82)90004-5}.

\bibitem[{Kalvelagen(2002)}]{kalvelagen2002}
Kalvelagen E (2002).
\newblock \enquote{Efficiently Solving DEA Models with \proglang{GAMS}.}
\newblock \emph{GAMS Development Corporation, Washington, DC}.
\newblock \urlprefix\url{http://amsterdamoptimization.com/pdf/dea.pdf}.

\bibitem[{Keshvari and Kuosmanen(2013)}]{Keshvari2013}
Keshvari A, Kuosmanen T (2013).
\newblock \enquote{{Stochastic Non-convex Envelopment of Data: Applying
  Isotonic Regression to Frontier Estimation}.}
\newblock \emph{European Journal of Operational Research}, \textbf{231},
  481--491.
\newblock \doi{10.1016/j.ejor.2013.06.005}.

\bibitem[{Koenker(2005)}]{Koenker2005b}
Koenker R (2005).
\newblock \emph{{Quantile Regression}}.
\newblock Cambridge University Press, Cambridge.
\newblock \doi{10.1017/CBO9780511754098}.

\bibitem[{Koenker and Bassett(1978)}]{Koenker1978}
Koenker R, Bassett G (1978).
\newblock \enquote{{Regression Quantiles}.}
\newblock \emph{Econometrica}, \textbf{46}, 33--50.
\newblock \doi{10.2307/1913643}.

\bibitem[{Kumbhakar and Lovell(2000)}]{Kumbhakar2000}
Kumbhakar SC, Lovell CK (2000).
\newblock \emph{{Stochastic Frontier Analysis}}.
\newblock Cambridge University Press, Cambridge.
\newblock \doi{10.1017/CBO9781139174411}.

\bibitem[{Kuosmanen(2006)}]{Kuosmanen2006}
Kuosmanen T (2006).
\newblock \enquote{{Stochastic Nonparametric Envelopment of Data: Combining
  Virtues of SFA and DEA in a Unified Framework}.}
\newblock \emph{{MTT Discussion Paper No.3/2006}}.

\bibitem[{Kuosmanen(2008)}]{Kuosmanen2008}
Kuosmanen T (2008).
\newblock \enquote{{Representation Theorem for Convex Nonparametric Least
  Squares}.}
\newblock \emph{Econometrics Journal}, \textbf{11}, 308--325.
\newblock \doi{10.1111/j.1368-423X.2008.00239.x}.

\bibitem[{Kuosmanen(2012)}]{Kuosmanen2012b}
Kuosmanen T (2012).
\newblock \enquote{{Stochastic Semi-Nonparametric Frontier Estimation of
  Electricity Distribution Networks: Application of the StoNED method in the
  Finnish regulatory model}.}
\newblock \emph{Energy Economics}, \textbf{34}, 2189--2199.
\newblock \doi{10.1016/j.eneco.2012.03.005}.

\bibitem[{Kuosmanen \emph{et~al.}(2015)Kuosmanen, Johnson, and
  Saastamoinen}]{Kuosmanen2015d}
Kuosmanen T, Johnson A, Saastamoinen A (2015).
\newblock \enquote{{Stochastic Nonparametric Approach to Efficiency Analysis: A
  Unified Framework}.}
\newblock In J~Zhu (ed.), \emph{Data Envelopment Analysis}, chapter~7, pp.
  191--244. Springer.

\bibitem[{Kuosmanen and Johnson(2010)}]{Kuosmanen2010a}
Kuosmanen T, Johnson AL (2010).
\newblock \enquote{{Data Envelopment Analysis as Nonparametric Least-Squares
  Regression}.}
\newblock \emph{Operations Research}, \textbf{58}, 149--160.
\newblock \doi{10.1287/opre.1090.0722}.

\bibitem[{Kuosmanen and Johnson(2017)}]{Kuosmanen2017a}
Kuosmanen T, Johnson AL (2017).
\newblock \enquote{{Modeling Joint Production of Multiple Outputs in StoNED:
  Directional Distance Function Approach}.}
\newblock \emph{European Journal of Operational Research}, \textbf{262},
  792--801.
\newblock \doi{10.1016/j.ejor.2017.04.014}.

\bibitem[{Kuosmanen and Kortelainen(2012)}]{Kuosmanen2012c}
Kuosmanen T, Kortelainen M (2012).
\newblock \enquote{{Stochastic Non-smooth Envelopment of Data: Semi-parametric
  Frontier Estimation Subject to Shape Constraints}.}
\newblock \emph{Journal of Productivity Analysis}, \textbf{38}, 11--28.
\newblock \doi{10.1007/s11123-010-0201-3}.

\bibitem[{Kuosmanen \emph{et~al.}(2013)Kuosmanen, Saastamoinen, and
  Sipil{\"{a}}inen}]{Kuosmanen2013}
Kuosmanen T, Saastamoinen A, Sipil{\"{a}}inen T (2013).
\newblock \enquote{{What is the Best Practice for Benchmark Regulation of
  Electricity Distribution? Comparison of DEA, SFA and StoNED Methods}.}
\newblock \emph{Energy Policy}, \textbf{61}, 740--750.
\newblock \doi{10.1016/j.enpol.2013.05.091}.

\bibitem[{Kuosmanen \emph{et~al.}(2021)Kuosmanen, Tan, and
  Dai}]{Kuosmanen2021a}
Kuosmanen T, Tan A, Dai S (2021).
\newblock \enquote{Performance of {English NHS} Hospitals During the First and
  Second Waves of the {COVID-19} Pandemic.}
\newblock \emph{working paper}.
\newblock Available online at
  https://www.researchgate.net/publication/349075148.

\bibitem[{Kuosmanen and Zhou(2021)}]{Kuosmanen2021}
Kuosmanen T, Zhou X (2021).
\newblock \enquote{Shadow Prices and Marginal Abatement Costs: Convex Quantile
  Regression Approach.}
\newblock \emph{European Journal of Operational Research}, \textbf{289}(2),
  666--675.
\newblock \doi{10.1016/j.ejor.2020.07.036}.

\bibitem[{Kuosmanen \emph{et~al.}(2020)Kuosmanen, Zhou, and
  Dai}]{Kuosmanen2020}
Kuosmanen T, Zhou X, Dai S (2020).
\newblock \enquote{How Much Climate Policy Has Cost for OECD Countries?}
\newblock \emph{World Development}, \textbf{125}, 104681.
\newblock \doi{10.1016/j.worlddev.2019.104681}.

\bibitem[{Lee and Cai(2020)}]{Lee2020}
Lee CY, Cai JY (2020).
\newblock \enquote{{LASSO Variable Selection in Data Envelopment Analysis with
  Small Datasets}.}
\newblock \emph{Omega}, \textbf{91}, 102019.
\newblock ISSN 03050483.
\newblock \doi{10.1016/j.omega.2018.12.008}.

\bibitem[{Lee \emph{et~al.}(2013)Lee, Johnson, Moreno-Centeno, and
  Kuosmanen}]{Lee2013}
Lee CY, Johnson AL, Moreno-Centeno E, Kuosmanen T (2013).
\newblock \enquote{{A More Efficient Algorithm for Convex Nonparametric Least
  Squares}.}
\newblock \emph{European Journal of Operational Research}, \textbf{227},
  391--400.
\newblock \doi{10.1016/j.ejor.2012.11.054}.

\bibitem[{Lim and Glynn(2012)}]{Lim2012}
Lim E, Glynn PW (2012).
\newblock \enquote{{Consistency of multidimensional convex regression}.}
\newblock \emph{Operations Research}, \textbf{60}, 196--208.
\newblock \doi{10.1287/opre.1110.1007}.

\bibitem[{Magnani and Boyd(2009)}]{Magnani2009}
Magnani A, Boyd SP (2009).
\newblock \enquote{{Convex piecewise-linear fitting}.}
\newblock \emph{Optimization and Engineering}, \textbf{10}, 1--17.
\newblock \doi{10.1007/s11081-008-9045-3}.

\bibitem[{Mazumder \emph{et~al.}(2019)Mazumder, Choudhury, Iyengar, and
  Sen}]{Mazumder2019}
Mazumder R, Choudhury A, Iyengar G, Sen B (2019).
\newblock \enquote{{A computational framework for multivariate convex
  regression and its variants}.}
\newblock \emph{Journal of the American Statistical Association}, \textbf{114},
  318--331.
\newblock \doi{10.1080/01621459.2017.1407771}.

\bibitem[{Meeusen \emph{et~al.}(1977)Meeusen, Van, and Broeck}]{Meeusen1977}
Meeusen W, Van J, Broeck D (1977).
\newblock \enquote{{Efficiency Estimation from Cobb-Douglas Production
  Functions with Composed Error}.}
\newblock \emph{International Economic Review}, \textbf{18}, 435--444.
\newblock \doi{10.2307/2525757}.

\bibitem[{{MOSEK ApS}(2021)}]{mosek}
{MOSEK ApS} (2021).
\newblock \emph{{The \pkg{MOSEK} Optimization Toolbox for \proglang{Python}
  Manual. Version 9.2.47}}.
\newblock \urlprefix\url{https://docs.mosek.com/9.2/toolbox/index.html}.

\bibitem[{Murtagh and Saunders(2003)}]{murtagh2012}
Murtagh B, Saunders M (2003).
\newblock \enquote{\pkg{MINOS} 5.51 User’s Guide.}
\newblock \urlprefix\url{http://stanford.edu/group/SOL/guides/minos551.pdf}.

\bibitem[{{Oh, DH and Suh, D}(2013)}]{oh2013}
{Oh, DH and Suh, D} (2013).
\newblock \enquote{{\pkg{nonparaeff}: Nonparametric Methods for Measuring
  Efficiency and Productivity}.}
\newblock \proglang{R} package version 0.5-8,
  \urlprefix\url{https://CRAN.R-project.org/package=nonparaeff}.

\bibitem[{{R Core Team}(2020)}]{R}
{R Core Team} (2020).
\newblock \emph{{\proglang{R}: A Language and Environment for Statistical
  Computing}}.
\newblock \proglang{R} Foundation for Statistical Computing, Vienna, Austria.
\newblock \urlprefix\url{https://www.R-project.org}.

\bibitem[{Raith \emph{et~al.}(2021)Raith, Perederieieva, Fauzi, Harton, Lee,
  Lin, Priddey, and Rouse}]{pyDEA}
Raith A, Perederieieva O, Fauzi F, Harton K, Lee A, Lin K, Priddey H, Rouse M
  (2021).
\newblock \enquote{\pkg{pyDEA}: Package for Conducting Data Envelopment
  Analysis.}
\newblock \proglang{Python} package version 1.6,
  \urlprefix\url{https://pypi.org/project/pyDEA}.

\bibitem[{Seijo and Sen(2011)}]{Seijo2011}
Seijo E, Sen B (2011).
\newblock \enquote{{Nonparametric least squares estimation of a multivariate
  convex regression function}.}
\newblock \emph{Annals of Statistics}, \textbf{39}, 1633--1657.
\newblock \doi{10.1214/10-AOS852}.

\bibitem[{Simm and Besstremyannaya(2020)}]{simm2020}
Simm J, Besstremyannaya G (2020).
\newblock \enquote{{\pkg{rDEA}: Robust Data Envelopment Analysis (DEA) for
  \proglang{R}}.}
\newblock \proglang{R} package version 1.2-6,
  \urlprefix\url{https://CRAN.R-project.org/package=rDEA}.

\bibitem[{Straub(2014)}]{ariane2014}
Straub A (2014).
\newblock \enquote{{\pkg{sfa}: Stochastic Frontier Analysis}.}
\newblock \proglang{R} package version 1.0-1,
  \urlprefix\url{https://CRAN.R-project.org/package=sfa}.

\bibitem[{{The Mathworks, Inc.}(2017)}]{MATLAB}
{The Mathworks, Inc} (2017).
\newblock \emph{{\proglang{MATLAB} — The Language of Technical Computing,
  Version R2015b}}.
\newblock Natick, Massachusetts.
\newblock \urlprefix\url{http://www.mathworks.com/products/matlab}.

\bibitem[{{The pandas development team}(2020)}]{reback2020}
{The pandas development team} (2020).
\newblock \enquote{Pandas-dev/pandas: \pkg{Pandas}.}
\newblock \doi{10.5281/zenodo.3509134}.

\bibitem[{Varian(1982)}]{Varian1982}
Varian HR (1982).
\newblock \enquote{{The Nonparametric Approach to Demand Analysis}.}
\newblock \emph{Econometrica}, \textbf{50}, 945.
\newblock \doi{10.2307/1912771}.

\bibitem[{Varian(1984)}]{Varian1984}
Varian HR (1984).
\newblock \enquote{{The Nonparametric Approach to Production Analysis}.}
\newblock \emph{Econometrica}, \textbf{52}, 579.
\newblock \doi{10.2307/1913466}.

\bibitem[{Virtanen \emph{et~al.}(2020)Virtanen, Gommers, Oliphant, Haberland,
  Reddy, Cournapeau, Burovski, Peterson, Weckesser, Bright, {van der Walt},
  Brett, Wilson, Millman, Mayorov, Nelson, Jones, Kern, Larson, Carey, Polat,
  Feng, Moore, {VanderPlas}, Laxalde, Perktold, Cimrman, Henriksen, Quintero,
  Harris, Archibald, Ribeiro, Pedregosa, {van Mulbregt}, and {SciPy 1.0
  Contributors}}]{2020SciPy-NMeth}
Virtanen P, Gommers R, Oliphant TE, Haberland M, Reddy T, Cournapeau D,
  Burovski E, Peterson P, Weckesser W, Bright J, {van der Walt} SJ, Brett M,
  Wilson J, Millman KJ, Mayorov N, Nelson ARJ, Jones E, Kern R, Larson E, Carey
  CJ, Polat {\.I}, Feng Y, Moore EW, {VanderPlas} J, Laxalde D, Perktold J,
  Cimrman R, Henriksen I, Quintero EA, Harris CR, Archibald AM, Ribeiro AH,
  Pedregosa F, {van Mulbregt} P, {SciPy 10 Contributors} (2020).
\newblock \enquote{{\pkg{SciPy} 1.0: Fundamental Algorithms for Scientific
  Computing in \proglang{Python}}.}
\newblock \emph{Nature Methods}, \textbf{17}, 261--272.
\newblock \doi{10.1038/s41592-019-0686-2}.

\bibitem[{Wang \emph{et~al.}(2014)Wang, Wang, Dang, and Ge}]{Wang2014c}
Wang Y, Wang S, Dang C, Ge W (2014).
\newblock \enquote{{Nonparametric Quantile Frontier Estimation Under Shape
  Restriction}.}
\newblock \emph{European Journal of Operational Research}, \textbf{232},
  671--678.
\newblock \doi{10.1016/j.ejor.2013.06.049}.

\bibitem[{Wilson(2008)}]{wilson2008}
Wilson PW (2008).
\newblock \enquote{{\pkg{FEAR}: A Software Package for Frontier Efficiency
  Analysis With \proglang{R}}.}
\newblock \emph{Socio-economic Planning Sciences}, \textbf{42}, 247--254.
\newblock \doi{10.1016/j.seps.2007.02.001}.

\bibitem[{Yagi \emph{et~al.}(2020)Yagi, Chen, Johnson, and
  Kuosmanen}]{Yagi2018}
Yagi D, Chen Y, Johnson AL, Kuosmanen T (2020).
\newblock \enquote{{Shape-Constrained Kernel-Weighted Least Squares: Estimating
  Production Functions for Chilean Manufacturing Industries}.}
\newblock \emph{Journal of Business and Economic Statistics}, \textbf{38},
  43--54.
\newblock \doi{10.1080/07350015.2018.1431128}.

\end{thebibliography}

\newpage
\begin{appendix}

\section{List of Acronyms} \label{app:acrony}
\vspace{-1em}
\begin{itemize}[align=left,labelwidth=\widthof{StoNEZD}, leftmargin=2.5cm]
  \setlength{\itemsep}{0pt}
  \setlength{\parskip}{0pt}
  \setlength{\parsep}{0pt}
  \item[COLS]: Corrected Ordinary Least Squares \par
  \item[C$^2$NLS]: \vspace{\fill} Corrected Convex Nonparametric Least Squares \par
  \item[CER]: Convex Expectile Regression \par
  \item[CNLS]: Convex Nonparametric Least Squares \par
  \item[CNLS-G]: Convex Nonparametric Least Squares Generic Algorithm  \par
  \item[CQR]: Convex Quantile Regression \par
  \item[CRS]: Constant Returns to Scale \par
  \item[DDF]: Directional Distance Function \par
  \item[DEA]: Data Envelopment Analysis \par
  \item[DMU]: Decision-Making Unit \par
  \item[FDH]: Free Disposal Hull \par
  \item[MOM]: Method of Moments \par
  \item[QLE]: Quasi-likelihood Estimation \par
  \item[ICNLS]: Isotonic Convex Nonparametric Least Square \par
  \item[ICER]: Isotonic Convex Quantile Regression \par
  \item[ICQR]: Isotonic Convex Expectile Regression \par
  \item[SFA]: Stochastic Frontier Analysis \par
  \item[StoNED]: Stochastic Nonparametric Envelopment of Data \par
  \item[StoNEZD]: Stochastic Semi-nonparametric Envelopment of Z Variables Data \par
  \item[KDE]: Kernel Density Estimation \par
  \item[VRS]: Variable Returns to Scale \par
\end{itemize}

\section{Estimated CNLS residuals: an additive VRS model (GAMS) } \label{app:est_res_gams}
\vspace{-1em}
\begin{CodeOutput}
----    160 VARIABLE e.L  error terms
1    -2.802,    2     1.414,    3   -22.224,    4  -350.911,    5   -13.700
6   101.015,    7   -28.872,    8   -14.040,    9    -0.847,    10   56.894
11  285.507,    12  679.384,    13  -20.230,    14  -70.008,    15   10.507
16   74.597,    17   -6.539,    18  -30.076,    19  -40.135,    20  -27.778
21   48.760,    22   87.008,    23   22.481,    24   23.951,    25   -2.142
26  -22.837,    27  -37.861,    28 -351.071,    29   66.858,    30  -21.703
31   -8.951,    32  216.006,    33   37.597,    34  -31.811,    35  -23.785
36 -201.341,    37  -69.290,    38    6.269,    39    3.756,    40  -74.607
41   -1.975,    42  -11.971,    43 -236.746,    44    6.575,    45   11.657
46  -20.004,    47  -67.215,    48   -5.239,    49  -55.947,    50  265.514
51    1.924,    52  -17.651,    53    4.656,    54   38.854,    55   -2.903
56  349.528,    57  163.991,    58   35.001,    59   28.393,    60  -99.132
61   32.789,    62 -604.040,    63  197.101,    64   21.411,    65  -26.755
66  -15.130,    67   29.384,    68 -128.271,    69  -16.445,    70  -13.386
71 -293.862,    72   -2.551,    73  476.620,    74   -9.804,    75   32.090
76  -17.241,    77  114.768,    78   -3.852,    79   35.396,    80  -62.937
81    8.919,    82  349.397,    83  -80.700,    84 -604.758,    85  -59.840
86    6.718,    87    6.174,    88   -6.061,    89   -0.885
\end{CodeOutput}

\newpage
\section{Estimated CNLS residuals: an additive VRS model (pyStoNED)} \label{app:est_res_py}
\vspace{-1em}
\begin{CodeOutput}
epsilon : residual
    Size=89, Index=I
    Key : Lower : Value               : Upper : Fixed : Stale : Domain
      0 :  None : -2.8024040090178914 :  None : False : False :  Reals
      1 :  None :  1.4140528128759229 :  None : False : False :  Reals
      2 :  None : -22.223744892648355 :  None : False : False :  Reals
      3 :  None :    -350.91097508901 :  None : False : False :  Reals
      4 :  None : -13.699899937048826 :  None : False : False :  Reals
      5 :  None :  101.01484316190096 :  None : False : False :  Reals
      6 :  None : -28.872348336757398 :  None : False : False :  Reals
      7 :  None : -14.039749134653078 :  None : False : False :  Reals
      8 :  None : -0.8474521090027167 :  None : False : False :  Reals
      9 :  None :   56.89430545783776 :  None : False : False :  Reals
     10 :  None :   285.5068982735597 :  None : False : False :  Reals
     11 :  None :   679.3838738157001 :  None : False : False :  Reals
     12 :  None : -20.230026149875016 :  None : False : False :  Reals
     13 :  None :  -70.00793999792549 :  None : False : False :  Reals
     14 :  None :  10.506530291158015 :  None : False : False :  Reals
     15 :  None :   74.59733583012911 :  None : False : False :  Reals
     16 :  None :  -6.538916384385139 :  None : False : False :  Reals
     17 :  None : -30.076412673918952 :  None : False : False :  Reals
     18 :  None :  -40.13473761433872 :  None : False : False :  Reals
     19 :  None : -27.777867439090812 :  None : False : False :  Reals
     20 :  None :   48.75945800794108 :  None : False : False :  Reals
     21 :  None :   87.00800578768963 :  None : False : False :  Reals
     22 :  None :  22.480756177157602 :  None : False : False :  Reals
     23 :  None :    23.9507283965699 :  None : False : False :  Reals
     24 :  None : -2.1416654007527427 :  None : False : False :  Reals
     25 :  None : -22.836604777669237 :  None : False : False :  Reals
     26 :  None : -37.861048214358334 :  None : False : False :  Reals
     27 :  None :  -351.0709400239564 :  None : False : False :  Reals
     28 :  None :    66.8575905653208 :  None : False : False :  Reals
     29 :  None : -21.702814864843447 :  None : False : False :  Reals
     30 :  None :  -8.951285786354433 :  None : False : False :  Reals
     31 :  None :   216.0059312936645 :  None : False : False :  Reals
     32 :  None :  37.597340051237836 :  None : False : False :  Reals
     33 :  None : -31.811496011595466 :  None : False : False :  Reals
     34 :  None : -23.784708760965316 :  None : False : False :  Reals
     35 :  None : -201.34125864236086 :  None : False : False :  Reals
     36 :  None :  -69.29033394425426 :  None : False : False :  Reals
     37 :  None :  6.2687614525916615 :  None : False : False :  Reals
     38 :  None :   3.755841883998798 :  None : False : False :  Reals
     39 :  None :  -74.60739790684079 :  None : False : False :  Reals
     40 :  None :  -1.974792422660812 :  None : False : False :  Reals
     41 :  None : -11.970877239816076 :  None : False : False :  Reals
     42 :  None :  -236.7462774012531 :  None : False : False :  Reals
     43 :  None :   6.574582643949043 :  None : False : False :  Reals
     44 :  None :  11.656881908592197 :  None : False : False :  Reals
     45 :  None : -20.003997143376267 :  None : False : False :  Reals
     46 :  None :  -67.21455740523754 :  None : False : False :  Reals
     47 :  None :    -5.2391132374253 :  None : False : False :  Reals
     48 :  None :  -55.94694781186843 :  None : False : False :  Reals
     49 :  None :  265.51412810307147 :  None : False : False :  Reals
     50 :  None :  1.9241988681286273 :  None : False : False :  Reals
     51 :  None : -17.650818142520706 :  None : False : False :  Reals
     52 :  None :   4.656303515995944 :  None : False : False :  Reals
     53 :  None :   38.85434607338044 :  None : False : False :  Reals
     54 :  None : -2.9033063476584573 :  None : False : False :  Reals
     55 :  None :   349.5282971070195 :  None : False : False :  Reals
     56 :  None :  163.99050232155884 :  None : False : False :  Reals
     57 :  None :   35.00136220662897 :  None : False : False :  Reals
     58 :  None :  28.392936012634948 :  None : False : False :  Reals
     59 :  None :  -99.13217033642508 :  None : False : False :  Reals
     60 :  None :   32.78887118253597 :  None : False : False :  Reals
     61 :  None :   -604.039828622092 :  None : False : False :  Reals
     62 :  None :  197.10103085237904 :  None : False : False :  Reals
     63 :  None :   21.41105729756117 :  None : False : False :  Reals
     64 :  None : -26.755385499560376 :  None : False : False :  Reals
     65 :  None : -15.129727585118061 :  None : False : False :  Reals
     66 :  None :  29.384158404973107 :  None : False : False :  Reals
     67 :  None : -128.27098971868338 :  None : False : False :  Reals
     68 :  None :  -16.44516975139834 :  None : False : False :  Reals
     69 :  None : -13.386282784154389 :  None : False : False :  Reals
     70 :  None :  -293.8615590996965 :  None : False : False :  Reals
     71 :  None : -2.5506262260430503 :  None : False : False :  Reals
     72 :  None :  476.61961342030463 :  None : False : False :  Reals
     73 :  None :  -9.803554843610236 :  None : False : False :  Reals
     74 :  None :  32.089472403185525 :  None : False : False :  Reals
     75 :  None :  -17.24135562459952 :  None : False : False :  Reals
     76 :  None :  114.76795801536423 :  None : False : False :  Reals
     77 :  None :  -3.851436745667538 :  None : False : False :  Reals
     78 :  None :   35.39622164193156 :  None : False : False :  Reals
     79 :  None :  -62.93737365034673 :  None : False : False :  Reals
     80 :  None :    8.91934740692409 :  None : False : False :  Reals
     81 :  None :   349.3972830690898 :  None : False : False :  Reals
     82 :  None :  -80.69964425322985 :  None : False : False :  Reals
     83 :  None :   -604.758144376533 :  None : False : False :  Reals
     84 :  None :  -59.83972941268223 :  None : False : False :  Reals
     85 :  None :  6.7184004638272405 :  None : False : False :  Reals
     86 :  None :   6.173694361311188 :  None : False : False :  Reals
     87 :  None :  -6.061039431942607 :  None : False : False :  Reals
     88 :  None :  -0.885163858119256 :  None : False : False :  Reals
\end{CodeOutput}

\end{appendix}

\end{document}